\documentclass[12pt]{article}

\usepackage{amssymb}
\usepackage{amssymb}
\usepackage{tikz}
\usetikzlibrary{arrows, calc, decorations.pathmorphing, backgrounds, positioning, fit, petri, automata}
\definecolor{grey1}{rgb}{0.5,0.5,0.5}
\usepackage[top=2.54cm, bottom=2.54cm, left=2.2cm, right=2.2cm]{geometry}
\usepackage{algorithmicx}
\usepackage{algcompatible}
\usepackage{algorithm}

\usepackage{changebar}

\usepackage{hyperref}
\hypersetup{ 
  colorlinks,
  citecolor=blue,
  linkcolor=red,
  urlcolor=magenta}

\usepackage{mathtools}
\usepackage{graphicx}
\usepackage{subfigure}
\usepackage{amsmath,amsthm,bm,bbm}
\usepackage{rotating}
\usepackage{mathrsfs}
\usepackage{chngcntr}
\usepackage{apptools}
\usepackage[titletoc,title]{appendix}
\usepackage{xcolor}         
\definecolor{grau}{rgb}{0.8,0.8,0.8}

\newcommand{\chen}[1]{\color{orange}}
\usepackage{setspace,lscape,longtable}
\usepackage{amsmath,amsthm,bm}
\usepackage{fullpage}  

\usepackage{titlesec}
\usepackage[english]{babel}
\usepackage[shortlabels]{enumitem}
\usepackage[comma,authoryear]{natbib}

\DeclareMathOperator*{\argmax}{arg\,max}

\newcommand{\prob}{{\mathbb{P}}}

\newcommand{\expect}{\mathbb{E}}

\newcommand{\iidsim}{{\overset{\mathrm{i.i.d.}}{\sim}}}
\newcommand{\transpose}{^{\mathrm{T}}}

\newcommand{\calD}{{\mathcal{D}}}

\newcommand{\calI}{{\mathcal{I}}}

\newcommand{\calL}{{\mathcal{L}}}

\newcommand{\calW}{{\mathcal{W}}}

\newcommand{\balpha}{{\boldsymbol{\alpha}}}

\newcommand{\bU}{{\mathbf{U}}}

\newcommand{\bY}{{\mathbf{Y}}}
\newcommand{\bv}{{\mathbf{v}}}
\newcommand{\bx}{{\mathbf{x}}}
\newcommand{\bX}{{\mathbf{X}}}

\newcommand{\bB}{{\mathbf{B}}}

\newcommand{\bD}{{\mathbf{D}}}

\newcommand{\bF}{{\mathbf{F}}}

\newcommand{\bM}{{\mathbf{M}}}

\newcommand{\bR}{{\mathbf{R}}}

\newcommand{\bZ}{{\mathbf{Z}}}
\newcommand{\bw}{{\mathbf{w}}}

\newcommand{\bu}{{\mathbf{u}}}

\newcommand{\by}{{\mathbf{y}}}
\newcommand{\bz}{{\mathbf{z}}}

\newcommand{\bb}{{\mathbf{b}}}

\newcommand{\bV}{{\mathbf{V}}}
\newcommand{\bgamma}{{\bm{\gamma}}}

\newcommand{\bbeta}{{\bm{\beta}}}
\newcommand{\bDelta}{{\bm{\Delta}}}

\newcommand{\bsigma}{{\bm{\sigma}}}

\newcommand{\eye}{{\mathbf{I}}}
\newcommand{\one}{{\mathbf{1}}}
\newcommand{\btheta}{{\bm{\theta}}}
\newcommand{\bTheta}{{\bm{\Theta}}}

\newcommand{\bmu}{{\bm{\mu}}}

\newcommand{\bPhi}{{\bm{\Phi}}}
\newcommand{\bPsi}{{\bm{\Psi}}}

\newcommand{\zero}{{\bm{0}}}
\newcommand{\eps}{\epsilon}

\newcommand{\keywords}[1]{\par\addvspace\baselineskip\noindent\enspace\ignorespaces \textbf{Keywords: }#1}

\title{Variational Bayesian Multiple Imputation in High-Dimensional Regression Models With Missing Responses}

\author{Qiushuang Li\thanks{Department of Epidemiology and Biostatistics, Temple University, 
1101 W. Montgomery Ave.
 Philadelphia, PA 19122} \and Recai M. Yucel\footnotemark[1]
 \thanks{Correspondence should be addressed to \href{mailto:recai.yucel@temple.edu}{recai.yucel@temple.edu} }
 }
 \date{}
\begin{document}
\maketitle
\begin{abstract}
Multiple imputation has become one of the standard methods in drawing inferences in many incomplete data applications.  Applications of multiple imputation in relatively more complex settings, such as high-dimensional clustered data, require specialized methods to overcome the computational burden. Using linear mixed-effects models, we develop such methods that can be applied to continuous, binary, or categorical incomplete data by employing variational Bayesian inference to sample the posterior predictive distribution of the missing data. These methods specifically target high-dimensional data and work with the spike-and-slab prior, which automatically selects the variables of importance to be in the imputation model. The individual regression computation is then incorporated into a variable-by-variable imputation algorithm. Finally, we use a calibration-based algorithm to adopt these methods to multiple imputations of categorical variables.  We present a simulation study and illustrate on National Survey of Children's Health data to assess the performance of these methods in a repetitive sampling framework. 
\end{abstract}


\subsection*{Statement of significance}
We propose a novel and computationally efficient algorithm for sequential hierarchical regression imputation using optimization-based variational Bayesian inference. Individual regression models utilize Bayesian linear mixed-effect models and allow a variable selection structure through the spike-and-slab prior on the fixed-effect regression coefficients. The optimization nature of the algorithm bypasses computational burden caused by the Markov chain Monte Carlo sampler. The proposed algorithm can also handle mixture of  continuous and categorical variables through the integration of a  calibration-based rounding strategy. 

\section{Introduction} 
\label{sec:introduction}
Missing data are typically norm rather than exception in a wide range of areas ranging from survey data analysis to  signal processing, compressed sensing \citep{candes2009exact,5452187,5454406,5714248}, collaborative filtering, and recommendation systems \citep{koren2009matrix}. For example, in the Netflix Prize competition, some movies that are not rated can be treated as missing. The participants must predict grades on the entire qualifying set with the scores for half of the data. The missing data problems also occur in computer experiments and biomedical applications because of equipment limitations \citep{bayarri2007framework}. 

To deal with missing data, statisticians have relied on various corrective methods including imputation methods, which can adversely impact the  statistical uncertainty if the imputed values are treated as if they were observed. Inference by multiple imputation (MI) aims to solve this aspect while also allowing users to proceed with the complete-data methods.  MI simply samples missing data from their underlying predictive distribution, and by doing so, statisticians aim to account for the uncertainty due to missing values in contrast to a single imputation \citep{rubinmultiple}. Implementing MI within a fully Bayesian model incorporates the uncertainty in the unknown parameters, which conventionally uses Markov Chain Monte Carlo (MCMC) techniques for computation.

In this work, we are particularly interested in incorporating variable selection routines in linear mixed-effect models into the MI inference for computational efficiency. Classic variable selection methods typically utilizes certain information-criterion-based model selection approaches, such as Akaike information criterion (AIC) \citep{1100705} and Bayesian information criterion (BIC) \citep{schwarz1978}. These model selection approaches suffer from computation inefficiency because the model complexity is exponential in the number of variables. The computation bottleneck of variable selection for linear models was successfully tackled by LASSO \citep{https://doi.org/10.1111/j.2517-6161.1996.tb02080.x} and its variants \citep{doi:10.1198/016214506000000735,https://doi.org/10.1111/j.1467-9868.2005.00503.x} through a collection of seminal convex optimization approach. 

Formulating and implementing these methods can be quite challenging from the perspective of missing data and MI. To address this issue, our recent work \cite{QiushuangLi2020} proposed to apply the spike-and-slab prior for variable selection in the context of a fully Bayesian model, such that simulation-based MI can be performed. Theoretical properties of the spike-and-slab prior for the standard linear regression model with high dimensionality have been explored in \cite{castillo2015bayesian}. However, as commented in \cite{castillo2015bayesian}, addressing the computational challenge for high-dimensional variable selection problems is out of the scope of fully Bayesian models at present due to the need to explore the entire space of all possible models, which is exponential in the number of variables. This problem becomes particularly challenging when MCMC is implemented. 

The main purpose of this work is to propose a computationally efficient method that can deal with high-dimensional variable selection problems in the linear mixed-effect model and, incorporate this into MI inference to deal with arbitrary missing data.
Our approach is based on an optimization-based approximate inference method called the \emph{variational Bayesian inference} \citep{bishop2006pattern}. Variational Bayesian inference methods have been developed for various statistical considerations some of which are latent variable models such as mixture models, stochastic block models, generalized linear models \citep{doi:10.1080/01621459.2018.1473776,han2019statistical}, nonparametric Bayesian models \citep{pati2018statistical,yang2020alpha,zhang2020convergence}, and high-dimensional statistical models \citep{doi:10.1080/01621459.2020.1847121,ning2021spike}.
In contrast to simulation-based MCMC methods with many potential convergence issues and complications, variational Bayesian inference is a collection of approximate Bayesian inference methods that are formulated as a mathematical optimization problem. Specifically, we develop a computationally efficient variational Bayesian inference algorithm for approximate inference for linear mixed-effect model with high-dimensional covariates. We further incorporate this into  sequential hierarchical regression imputation (SHRIMP) \citep{yucel2017sequential} for continuous data and the calibration-based imputation \citep{doi:10.1198/000313008X300912,doi:10.1002/sim.4355} for binary and categorical data. The advantage of the proposed method is that it addresses the computation bottleneck of variable selection problems in Bayesian models through approximate inference and allows MI inference to be conducted to deal with missing data. 

The rest of this working paper is arranged as follows. In Section \ref{sec:linear_mixed_effects_regression_models}, we first briefly review the high-dimensional linear mixed-effect model with missing responses and then elaborate on the Bayesian model with the spike-and-slab prior. In Section \ref{sub:variational_inference_with_spike_and_slab_}, we develop the proposed variational Bayesian inference algorithm with the spike-and-slab prior for variable selection in the presence of missing responses. Section \ref{sec:application_in_multiple_imputation_of_missing_data} demonstrates how the variational Bayesian inference algorithm serves as a building block that can be embedded for different MI methods for missing responses, including the sequential hierarchical regression imputation method for continuous data and the calibration-based routine for categorical data. Section \ref{sec:numerical_examples} presents a comprehensive simulation study to gauge the performance of our method and demonstrates its application in the National Survey of Children's Health. We conclude with Section \ref{sec:discussion} with a discussion and future directions.  

\noindent\textbf{Notation:} In a clustered data setup with $m$ clusters and the number of observations for the $i$th cluster being $n_i$, $i\in\{1,\ldots,m\}$, we let $y_{ij}$ denote the observation corresponding to the $j$th individual in the $i$th cluster and is subject to missingness, and $\bx_{ij}$ is the vector of the remaining variables that are fully observed (or has been imputed), $i\in\{1,\ldots,m\}$, $j\in\{1,\ldots,n_i\}$. We use $\bY_{(\mathrm{mis})}$ and $\bY_{(\mathrm{obs})}$ to denote the missing portion and observed portion of $\bY = (y_{ij}:i\in\{1,\ldots,m\},j\in\{1,\ldots,n_i\})$. Similarly, $\mathbb{X}_{(\mathrm{mis})}$ and $\mathbb{X}_{(\mathrm{obs})}$ denote the missing and observed portions of $\mathbb{X} = (\bx_{ij}:i\in\{1,\ldots,m\},j\in\{1,\ldots,n_i\})$. Let $R_{y_{ij}}$ denote the missingness indicator for $y_{ij}$, namely, $R_{y_{ij}} = 1$ if $y_{ij}$ is missing and $R_{y_{ij}} = 0$ if $y_{ij}$ is observed. We assume that the missingness mechanism throughout this work is missing at random (MAR), namely, $p(\bR_\by\mid\bX,\bY) = p(\bR_\by\mid\bX_{(\mathrm{obs})},\bY_{(\mathrm{obs})})$.

\section{High-Dimensional Linear Mixed-effect Model with Missing Responses} 
\label{sec:linear_mixed_effects_regression_models}


We assume a linear mixed-effects model for continuous response variable $y_{ij}$ \citep{random_effect_model}:
\begin{align}
	\label{eqn:LME_model}
	y_{ij} = \bx_{ij}\transpose\bbeta + \bZ_{ij}\transpose\bb_i + \eps_{ij},\quad i = 1,\ldots,m,\quad j = 1,\ldots,n_i,
\end{align}
where $\bbeta\in\mathbb{R}^p$ is the fixed-effects regression coefficient, $\bx_{ij}$'s are the fixed-effects covariates, $\bb_1,\ldots,\bb_m\iidsim\mathrm{N}(0,\bPsi)$ are the random effects, and $\eps_{ij}\iidsim\mathrm{N}(0, \sigma_e^2)$ are the errors. Here, $i\in\{1,\ldots,m\}$ is the index for clusters and $j\in\{1,\ldots,n_i\}$ denotes the $j$th subject in the $i$th cluster, where $n_i$ denotes the number of subjects in the $i$th cluster. The responses $y_{ij}$'s are either observed or missing, 
and $\bx_{ij}\in\mathbb{R}^p$'s are the individual-level covariates that can also be either observed or missing, and the missing portion can be imputed within our algorithm as described in Section \ref{sec:application_in_multiple_imputation_of_missing_data}. 

In this work, we consider a scenario where the number of covariates $p$ is comparable or even larger than the sample size. The fixed-effect regression coefficient vector $\bbeta$ is assumed to be sparse, i.e., the number of non-zero entries of $\bbeta$ is comparably smaller than the sample size. Consequently, the number of active covariates, namely, those entries of $\bx_{ij}$'s corresponding to the non-zero entries of $\bbeta$, is also significantly smaller than the sample size. Our main focus here is to sample the missing portion of the responses $y_{ij}$'s from their predictive distributions (hence multiple imputation inference) but also incorporate the variable selection structure due to the sparsity of $\bbeta$. Leveraging a fully Bayesian model, our recent work \cite{QiushuangLi2020} developed a Gibbs sampler to draw posterior samples from the joint distribution of $(\bbeta,\bb_1,\dots,\bb_m,\sigma_{\bb_i},\sigma_e)$ the missing data $\bY_{(\mathrm{mis})}$ from the corresponding posterior predictive distribution. To select the variables among $x_{ij1},\dots,x_{ijp}$, a spike-and-slab prior distribution is assigned to the regression coefficient $\bbeta$, and this spike-and-slab variable selection approach has been broadly applied to Bayesian variable selection. When we have to deal with the high dimensional model where the dimension $p$ of the regression coefficient vector $\bbeta$ is comparable or significantly larger than $m$ or $n$, the variable selection process in the Markov Chain Monte Carlo can be extremely slow because the algorithm requires randomly searching the model space with $2^p$ candidate submodels. In what follows, we develop a computationally efficient optimization-based variational Bayesian inference for variable selection in the presence of missing data.   

\subsection{Background on variational Bayesian inference} 
\label{sub:Variational Inference}
Variational Bayesian (VB) inference is a family of approximate Bayesian inference methods that differ from classical simulation-based Bayesian inference methods (e.g., MCMC or approximate Bayesian computation) (see \cite{bishop2006pattern} for more detailed description). Compared to Markov Chain Monte Carlo samplers, it is formulated as a mathematical optimization problem and is considerably faster than MCMC. VB begins with a fully Bayesian model by specifying prior distributions of the model unknowns. Unknowns here are random-effects, missing portion of the responses 
($\bY_{(\mathrm{mis})}$) and model parameters ($\bTheta$).  Let $\bPhi$ denote all these unknowns: 
\[
\bPhi=\{\bY_{(\mathrm{mis})}, \bbeta,\bb_1,\dots,\bb_n,\tau,\bPsi,\bV,\nu,w,\gamma,\sigma_e,\sigma_0,\mu_0\}.
\]
We first specify the complete Bayesian model through the prior distribution $p(\bTheta)$, and our goal is to find a distribution $q(\bPhi)$ as an approximation for the posterior distribution $p(\bPhi\mid\bY)$. The distribution $q$ is referred  as the {\it variational distribution}. Let us consider the following decomposition of the log marginal distribution of the observed responses $\ln p(\bY_{\mathrm{(obs)}})$:
\[
\ln p(\bY_{(\mathrm{obs})})=\calL(q)+\mathrm{KL}(q\|p),
\]
where $\calL(q)$ is referred to as the \emph{evidence lower bound} (ELBO) that can be written as
\begin{align*}
	\calL(q)&=\int q(\bPhi)\ln\left\{\frac{p(\bY,\bPhi)}{q(\bPhi)}\right\}\mathrm{d}\bPhi,
  \end{align*}
and $\mathrm{KL}(q\|p)$ is the Kullback-Leibler (KL) divergence between the variational distribution $q$ and the posterior distribution $p(\bPhi\mid\bY_{\mathrm{(obs)}})$ of $\bPhi$ given $\bY_{\mathrm{(obs)}}$:
\begin{align*}
	\mathrm{KL}(q\|p) = -\int q(\bPhi)\ln\left\{\frac{p(\bY,\bPhi)}{q(\bPhi)}\right\}\mathrm{d}\bPhi\nonumber.
\end{align*}
To obtain an approximation $q$ for the posterior distribution, a reasonable choice is to minimize the KL divergence $\mathrm{KL}(q\|p)$, which is equivalent to maximizing the ELBO $\calL(q)$. Note that the marginal distribution $p(\bY_{(\mathrm{obs})})$ is typically intractable to compute in practice, and the variational inference algorithm circumvents this complication by maximizing the more tractable ELBO objective function. 

\subsection{Bayesian linear mixed-effect model with a spike-and-slab prior} 
\label{sub:bayesian_linear_mixed_effect_model}


To lay the foundation of the variational Bayesian inference in the context of the linear mixed-effect model with missing responses, we first specify a fully Bayesian model by assigning a hierarchical prior distribution to $\bTheta$. First, note the following unknowns: 
\[
\{\bY_{\mathrm{(mis)}}, \bbeta,\bB,\sigma_e^2, \bPsi, \mu_0, \sigma_0^2, w, \gamma\},
\]
where $\bB = [\bb_1,\ldots,\bb_m]\in\mathbb{R}^{l\times m}$. Denote by $\calI_{(\mathrm{mis})} = \{(i, j):R_{y_{ij}} = 1,i = 1,\ldots,m,j = 1,\ldots,n_i\}$ the set of indices $(i, j)$ corresponding to $\bY_{(\mathrm{mis})}$ and $\calI_{(\mathrm{obs})}
 $ the indices corresponding to $\bY_{(\mathrm{obs})}$. The sampling model of the complete data $(\bY_{\mathrm{(obs)}}, \bY_{\mathrm{(mis)}}, \bB)$ can be described as
\begin{align}\label{eqn:complete_data_likelihood}
\begin{aligned}
p(\bY_{\mathrm{(obs)}}, \bY_{\mathrm{(mis)}}, \bZ\mid\bTheta) & = p(\bY_{\mathrm{(obs)}}\mid\bB, \bTheta)p(\bY_{\mathrm{(mis)}}\mid\bY_{\mathrm{(obs)}}, \bB, \bTheta)p(\bB\mid\bTheta),\\
p(\bY_{\mathrm{(obs)}}\mid\bB) & = \prod_{(i, j)\in\calI_{(\mathrm{obs})}}p(y_{ij}\mid\bbeta, \bb_i, \sigma_e^2),\\
p(\bY_{\mathrm{(mis)}}\mid\bY_{\mathrm{(obs)}},\bB) & = \prod_{(i, j)\in\calI_{(\mathrm{mis})}}p(y_{ij}\mid\bbeta, \bb_i, \sigma_e^2),\\
p(\bB\mid\bPsi) & = \prod_{i = 1}^np(\bb_i\mid\bPsi),
\end{aligned}
\end{align}
where
\begin{align*}
p(y_{ij}\mid \bbeta, \bb_i, \sigma_e^2)&=\mathrm{N}(y_{ij}\mid\bx_{ij}\transpose\bbeta+\bz_{ij}\transpose\bb_i,\sigma_e^2)=\frac{1}{\sqrt{2\pi\sigma_e^2}}\exp\left\{-\frac{(y_{ij}-\bx_{ij}\transpose\bbeta-\bz_{ij}\transpose\bb_i)^2}{2\sigma_e^2}\right\},\\
  p(\bb_i|\bPsi)&=\mathrm{N}(\bb_i\mid 0,\bPsi)=\prod_{i=1}^m\frac{1}{(\sqrt{2\pi})^l|\bPsi|^{\frac{1}{2}}}\exp\left(-\frac{1}{2}\bb_i\transpose\bPsi^{-1}\bb_i\right).
\end{align*}
The key to enforce sparsity in the fixed-effect regression coefficient vector $\bbeta$ lies in the spike-and-slab prior distribution \citep{castillo2015bayesian,castillo2012needles,variable_selection_spike_and_slab}. Specifically, for each entry $\beta_k$, $k = 1,\ldots,p$, the spike-and-slab distribution allows $\beta_k$ to take zero with a strictly positive probability $w$, and with probability $1 - w$, $\beta_k$ is generated from an absolutely continuous distribution supported on $\mathbb{R}$ (here we specifically take the continuous component to be a normal distribution). Formally, given the zero selection probability $w$, an auxiliary component assignment variable $\gamma_k$ is generated from $\mathrm{Bernoulli}(1 - w)$. The auxiliary variable $\gamma_k$ has the following interpretation: if $\gamma_k = 0$, then we set $\beta_k = 0$, and if $\gamma_k = 1$, then we draw $\beta_k$ from the continuous component of the spike-and-slab distribution. The prior samples $\beta_1,\ldots,\beta_p$ independently given the selection probability $w$. Consequently, the conditional prior of $\bbeta$ given $w$ can be described as follows:
\begin{align}\label{eqn:spike_and_slab}
\begin{aligned}
	p(\bbeta|\gamma_k,\mu_0,\sigma_0^2)\mathrm{d}\bbeta
  & = \prod_{k=1}^p\{\mathrm{N}(\bbeta_k|\mu_0,\sigma_0^2)\mathrm{d}\bbeta_k\}^{\gamma_k}\{\delta_0\mathrm{d}\bbeta_k\}^{1-\gamma_k}\\
  & = \prod_{k=1}^p\left[\frac{1}{\sqrt{2\pi\sigma_0^2}}\exp\left\{-\frac{1}{2\sigma_0^2}(\bbeta_k-\mu_0)^2\right\}\mathrm{d}\bbeta_k\right]^{\gamma_k}\left[\delta_0\mathrm{d}\bbeta_k\right]^{1-\gamma_k},\\
	p(\gamma_k|w)&=\prod_{k=1}^pw^{\gamma_k}(1-w)^{1-\gamma_k},
\end{aligned}
\end{align}
where $(\mu_0, \sigma_0^2)$ are the hyperparameters. The complete hierarchical prior distribution is completed by assigning the following hyperprior distributions to the hyperparameters of this specficiation, as well as $\sigma_e^2$ and $\bPsi$ for the sake of conjugacy:
\begin{align}\label{eqn:hyperpriors}
\begin{aligned}
	p(w)& = \mathrm{Beta}(w\mid a_w, b_w)\propto w^{a_w-1}(1-w)^{b_w-1},\\
	p(\sigma_e^2)& = \mathrm{IG}(\sigma_e^2\mid a_1, b_1) = \frac{b_1^{a_1}}{\Gamma(a_1)}\left(\frac{1}{\sigma_e^2}\right)^{a_1+1}\exp\left\{-\frac{b_1}{\sigma_e^2}\right\},\\
  p(\mu_0) & = \mathrm{N}(\mu_0\mid 0 , 1) = \frac{1}{\sqrt{2\pi}}e^{-\mu_0^2/2},\\
	p(\sigma_0^2)& = \mathrm{IG}(\sigma_0^2\mid 1, 1) = \left(\frac{1}{\sigma_0^2}\right)^{2}\exp\left\{-\frac{1}{\sigma_0^2}\right\},\\
	p(\bPsi^{-1})& = \calW(\bPsi^{-1}\mid\nu, \bV^{-1})\propto|\bPsi^{-1}|^{\frac{\nu-l-1}{2}}\exp\left\{-\frac{1}{2}\mathrm{tr}(\bPsi^{-1}\bV^{-1})\right\}.
\end{aligned}
\end{align}
This completes the specification of our hierarchical Bayesian model.

\section{Variational Bayesian Inference With a Spike-and-Slab Prior} 
\label{sub:variational_inference_with_spike_and_slab_}

We follow the commonly adopted mean-field approximation assumption and set the variational distributions in the following factorized form \citep{bishop2006pattern}:
\begin{align}
\label{eqn:mean_field_approximation}
q(\bPhi)
 = q(\bY_{(\mathrm{mis})})q(\bbeta,\gamma)q(\bB)q(\sigma_e^2)q(\bPsi)q(\mu_0)q(\sigma_0^2)q(w),
\end{align}
where $\bgamma = [\gamma_1,\ldots,\gamma_p]\transpose{}$ is the $p$-dimensional auxiliary variable specifying the sparsity of $\bbeta$ in the spike-and-slab distribution setup. Note, however, that the mathematical optimization problem
\begin{align*}
&\max_{q} \calL\{q(\bPhi)\}
\end{align*}
is an infinite-dimensional optimization problem because constraint \eqref{eqn:mean_field_approximation} is still an infinite-dimensional statistical manifold. Although an easy-to-implement coordinate-ascent variational algorithm can be obtained when the likelihood is in the exponential family form as suggested by \cite{bishop2006pattern}, our case brings additional computational challenge as we introduce the spike-and-slab prior \eqref{eqn:spike_and_slab} for $\bbeta$ with singularity. Alternatively, it is also reasonable to posit certain parametric forms of the variational distribution $q$ such that \eqref{eqn:mean_field_approximation} can be further reduced to a finite-dimensional statistical manifold (see, for example, \citealp{blei2003latent}). 
Hence, we further assume the following parametric form of the variational distribution for the sake of conjugacy:
\begin{align}\label{eqn:variational_distribution}
\begin{aligned}
	q(\bY_{(\mathrm{mis})}\mid (\widehat{\mu}_{y_{ij}}, \widehat{\sigma}_{y_{ij}}^2)_{(i, j)\in\calI_{(\mathrm{mis})}}) & = \prod_{(i,j)\in\calI_{(\mathrm{mis})}}\mathrm{N}(y_{ij}\mid \widehat{\mu}_{y_{ij}}, \widehat{\sigma}_{y_{ij}}^2),\\
    q(\bbeta,\gamma\mid (\theta_k, \widehat{\mu}_{\beta_k}, \widehat{\sigma}_{\beta_k}^2)_{k = 1}^p)\mathrm{d}\bbeta&=\prod_{k=1}^p\left\{\theta_k\mathrm{N}(\bbeta_k\mid\hat{\mu}_{\beta_k},\hat{\sigma}_{\beta_k}^2)\mathrm{d}\beta_k\right\}^{\gamma_k}\left\{(1-\theta_k)\delta_0\mathrm{d}\beta_k\right\}^{1-\gamma_k},\\
    q(\bB\mid\widehat{\bmu}_{\bb_1},\widehat{\bPsi}_1,\ldots,\widehat{\bmu}_{\bb_m}, \widehat{\bPsi}_m)&=\prod_{i=1}^m\mathrm{N}(\bb_i\mid\widehat{\bmu}_{\bb_i},\widehat{\bPsi}_{\bb_i}),
    \\
    q(\sigma_e^2\mid\hat{a}_{\sigma_e^2},\hat{b}_{\sigma_e^2})&=\mathrm{IG}(\sigma_e^2\mid\hat{a}_{\sigma_e^2},\hat{b}_{\sigma_e^2})
    ,\\
    q(\bPsi^{-1}\mid\widehat{\bPsi})\mathrm{d}\bPsi^{-1}&=\delta_{\widehat{\bPsi}^{-1}}(\mathrm{d}\bPsi^{-1})
    ,\\
    q(\mu_0\mid\hat{\mu}_{\mu_0},\hat{\sigma}_{\mu_0}^2)&=\mathrm{N}(\mu_0\mid\hat{\mu}_{\mu_0},\hat{\sigma}_{\mu_0}^2)
    ,\\
    q(\sigma_0^2\mid\hat{a}_{\sigma_0^2},\hat{b}_{\sigma_0^2})&=\mathrm{IG}(\sigma_0^2\mid\hat{a}_{\sigma_0^2},\hat{b}_{\sigma_0^2})
    ,\\
    q(w\mid\hat{a}_w,\hat{b}_w)&=\mathrm{Beta}(w\mid\hat{a}_w,\hat{b}_w)
    =
    \frac{w^{\hat{a}_w-1}(1-w)^{\hat{b}_w-1}}{B(\hat{a}_w,\hat{b}_w)}
    .
\end{aligned}
\end{align}
Note that we could use a Wishart variational distribution to approximate the posterior of the precision matrix $\bPsi^{-1}$ for the random effect coefficient $\bb_1,\ldots,\bb_m$. Here, we use a Dirac point mass at $\widehat{\bPsi}^{-1}$ to indicate that a point estimator for $\bPsi$ is taken, and the resulting solution is a maximum \emph{a posteriori} estimator for $\bPsi$. We also note that the set of parameters
\begin{align*}
\boldsymbol{\Xi} := \mathrel{\Big\{}&
(\widehat{\mu}_{y_{ij}}, \widehat{\sigma}_{y_{ij}}^2)_{i,j\in\calI_{(\mathrm{mis})}},(\theta_k, \widehat{\mu}_{\beta_k}, \widehat{\sigma}_{\beta_k}^2)_{k = 1}^p, (\widehat{\bmu}_{\bb_i}, \widehat{\bPsi}_{\bb_i})_{i = 1}^m, \\
    &(\widehat{a}_{\sigma_e^2}, \widehat{b}_{\sigma_e^2}), (\widehat{a}_{\sigma_e^2}, \widehat{b}_{\sigma_e^2}), 
    \widehat{\bPsi}, (\widehat{\mu}_{\mu_0}, \widehat{\sigma}_{\mu_0}^2),  (\widehat{a}_{w}, \widehat{b}_{w})
\mathrel{\Big\}}
\end{align*}
are the variables to be solved by optimizing the objective function $\calL(q)$. Then, variational Bayesian inference is implemented by iteratively maximizing the objective function $\calL(q)$ with respect to the variational parameters $\boldsymbol{\Xi}$. Below, we present the complete variational Bayesian inference algorithm. The detailed derivation is provided in the Appendix.  Note that here, we overwrite the sequence of iterates of the variational parameters when the algorithm proceeds and only keep track of the most updated values of these parameter values. For example, $\btheta\longleftarrow \bF(\btheta)$ means that we compute $\bF(\btheta)$ where $\btheta$ is the most recent value of the parameter $\btheta$, and then assign $\btheta$ with value $\bF(\btheta)$ so that the previous value of $\btheta$ is overwritten with the value $\bF(\btheta)$. This greatly simplifies the notations as well as releasing unnecessary computational memories. 
\begin{enumerate}
\item[$\blacksquare$] \textbf{Input: }
      
      Response data $\bY = (\by_{ij}:i\in\{1,\ldots,n\},j\in\{1,\ldots,n_i\})$ (with potentially missing entries)
      
      Fixed effect covariates $\mathbbm{X} = (\bx_{ij}:i\in\{1,\ldots,m\},j\in\{1,\ldots,n_i\})$
      
      Random effect covariates $\mathbbm{Z} = (\bz_{ij}:i\in\{1,\ldots,m\},j\in\{1,\ldots,n_i\})$

\item \textbf{Step 1:} Randomly initialize the variational parameters
      \begin{align*}
      \boldsymbol{\Xi} := \mathrel{\Big\{}&
(\widehat{\mu}_{y_{ij}}, \widehat{\sigma}_{y_{ij}}^2)_{i,j\in\calI_{(\mathrm{mis})}},(\theta_k, \widehat{\mu}_{\beta_k}, \widehat{\sigma}_{\beta_k}^2)_{k = 1}^p, (\widehat{\bmu}_{\bb_i}, \widehat{\bPsi}_{\bb_i})_{i = 1}^m, \\
    &(\widehat{a}_{\sigma_e^2}, \widehat{b}_{\sigma_e^2}), (\widehat{a}_{\sigma_e^2}, \widehat{b}_{\sigma_e^2}), 
    \widehat{\bPsi}, (\widehat{\mu}_{\mu_0}, \widehat{\sigma}_{\mu_0}^2),  (\widehat{a}_{w}, \widehat{b}_{w})
\mathrel{\Big\}}.
      \end{align*}
      and compute the following matrices related to the fixed-effect covariate $\mathbb{\bX}$:
      \[
      \bD_1(\mathbb{\bX}) = \sum_{i = 1}^m\sum_{j = 1}^{n_i}\bx_{ij}\bx_{ij}\transpose, \quad
      \bD_2(\mathbb{\bX}) = \sum_{i = 1}^m\sum_{j = 1}^{n_i}\mathrm{diag}(x_{ij1}^2,\ldots,x_{ijp}^2). 
      \]

\item \textbf{Step 2:} Loop until the objective function $\Omega$ converges (which occurs when $\Omega$ stabilizes and does not change much when the between the iterations):
      \begin{itemize}
        \item Update $\widehat\bPsi$ by maximizing $\Omega$ with respect to $\widehat\bPsi$: 
        This yields
        \[
        \widehat\bPsi \longleftarrow \frac{1}{(m + \nu - l - 1)}\left\{\sum_{i = 1}^m(\widehat{\bmu}_{\bb_i}\widehat{\bmu}_{\bb_i}\transpose + \widehat{\bPsi}_{\bb_i}) + \bV^{-1}\right\}
        \]
        \item Update $\widehat{\mu}_{y_{ij}}, \widehat{\sigma}^2_{y_{ij}}$ for all $(i, j)\in\calI_{(\mathrm{mis})}$
        using the following formula:
        \begin{align*}
        \widehat{\mu}_{ij} &\longleftarrow \sum_{k = 1}^px_{ijk}\theta_k\widehat{\mu}_{\beta_k} + \bz_{ij}\transpose\widehat{\bmu}_{\bb_i},
        \quad
        \widehat{\sigma}_{y_{ij}}^2 \longleftarrow \frac{\widehat{a}_{\sigma_e^2}}{\widehat{b}_{\sigma_e^2}}.
        \end{align*}
        \item Update $(\theta_k, \widehat{\mu}_{\beta_k}, \widehat{\sigma}_{\beta_k}^2)$ for all $k = 1,\ldots,p$. This step is the most complicated step of the entire algorithm, and we seek an updating rule that can be vectorized. For all $i = 1,\ldots,m$ and $j = 1,\ldots,n_i$, we set
        \begin{align*}
          \widehat{y}_{ij} & \longleftarrow \left\{
          \begin{aligned}
            &y_{ij},&\quad\text{if }(i, j)\notin\calI_{(\mathrm{mis})},\\
            &\widehat{\mu}_{y_{ij}},&\quad\text{if }(i, j)\in\calI_{(\mathrm{mis})},
          \end{aligned}
          \right.,
          \\
          \langle {y}_{ij}^2\rangle_q &\longleftarrow \left\{
          \begin{aligned}
            &y^2_{ij},&\quad\text{if }(i, j)\notin\calI_{(\mathrm{mis})},\\
            &\widehat{\mu}_{y_{ij}}^2 + \widehat{\sigma}_{y_{ij}}^2,&\quad\text{if }(i, j)\in\calI_{(\mathrm{mis})}.
          \end{aligned}
          \right.
          \end{align*}
          Set 
          \[
          \bv \longleftarrow \sum_{i = 1}^m\sum_{j = 1}^{n_i}\bx_{ij}(\widehat{y}_{ij} - \bz_{ij}\transpose{}\widehat{\bmu}_{\bb_i}). 
          \]
          Next, we consider vectorized updating formula for $\btheta, \widehat\bmu_\bbeta, \widehat\bsigma^2_\bbeta$, where $\btheta = [\theta_1,\ldots,\theta_p]\transpose{}$, $\widehat\bmu_\bbeta = [\widehat{\mu}_{\beta_1},\ldots,\widehat{\mu}_{\beta_p}]\transpose{}$, and $\widehat{\bsigma}_\bbeta^2 = [\widehat{\sigma}_{\beta_1}^2,\ldots,\widehat{\sigma}_{\beta_p}^2]\transpose$. Denote by $\one_p = [1,\ldots,1]\transpose{}\in\mathbb{R}^p$. For any function $f:\calD\subset\mathbb{R}\to\mathbb{R}$, we denote by $f(\bx)$ generically as the entrywise application of $f$ to the elements of $\bx$, namely, $f([x_1,\ldots,x_p]\transpose) := [f(x_1),\ldots,f(x_p)]\transpose$. Denote by $\bx_1\circ\bx_2$ as the entrywise product between two vectors $\bx_1,\bx_2$ of the same dimension, and $\bx_1/\bx_2$ as the entrywise ratio between $\bx_1$ and $\bx_2$. Finally, we use $\psi(\cdot)$ to denote the digamma function $\psi(x) = (\mathrm{d}/\mathrm{d}x)\ln[\Gamma(x)]$, where $\Gamma(\cdot)$ is the Gamma function. Then $\btheta$ can be obtained by first computing the update for $\mathrm{logit}(\btheta)$:
          \begin{align*}
          \mathrm{logit}(\btheta) & \longleftarrow
          \left[\psi(\widehat{a}_w) - \psi(\widehat{b}_w) + \frac{1}{2}\psi(\widehat{a}_{\sigma_0^2}) - \frac{1}{2}\ln(\widehat{b}_{\sigma_0^2})\right]\one_p + \frac{1}{2}\ln(\widehat{\bsigma}_{\bbeta}^2) + \frac{1}{2}\one_p\\
          &\quad\quad -\frac{1}{2}\left(\frac{\widehat{a}_{\sigma_0^2}}{\widehat{b}_{\sigma_0^2}}\right)[(\widehat{\bmu}_{\bbeta_k} - \widehat{\mu}_{\mu_0}\one_p)\circ(\widehat{\bmu}_{\bbeta_k} - \widehat{\mu}_{\mu_0}\one_p) + \widehat{\sigma}_{\mu_0}^2\one_p + \widehat{\bsigma}_{\bbeta}^2]\\
          &\quad\quad + \widehat{\bmu}_{\bbeta}^2\circ\left(\frac{1}{\widehat{\bsigma}_{\bbeta}^2}\right) - \left(\frac{\widehat{a}_{\sigma_0^2}}{\widehat{b}_{\sigma_0^2}}\right){\widehat{\bmu}_{\bbeta}}{\widehat{\mu}_{\mu_0}}
           - \left(\frac{\widehat{a}_{\sigma_0^2}}{\widehat{b}_{\sigma_0^2}}\right){\widehat{\bmu}_{\bbeta}}^2
          \\&\quad\quad
           - \frac{1}{2}\left(\frac{\widehat{a}_{\sigma_e^2}}{\widehat{b}_{\sigma_e^2}}\right)\sum_{i = 1}^m\sum_{j = 1}^n\bx_{ij}\circ\bx_{ij}\circ(\widehat{\bmu}_{\bbeta}\circ\widehat{\bmu}_{\bbeta} + \bsigma_{\bbeta}^2).
           \end{align*}
          Then $\btheta$ can be directly obtained by taking $\btheta\longleftarrow \mathrm{logit}^{-1}(\mathrm{logit}(\btheta))$. Set 
          \[
          \bTheta = \mathrm{diag}(\theta_1,\ldots, \theta_p).
          \]
          We then compute the updates for $\widehat{\bmu}_\bbeta$ and $\widehat{\bsigma}_\bbeta^2$: 
          \begin{align*}
          \widehat{\bsigma}_{\bbeta}^2&\longleftarrow
          \left(\frac{\widehat{a}_{\sigma_e^2}}{\widehat{b}_{\sigma_e^2}}\sum_{i = 1}^m\sum_{j = 1}^{n_i}\bx_{ij}\circ\bx_{ij} + \frac{\widehat{a}_{\sigma_0^2}}{\widehat{b}_{\sigma_0^2}}\one_p\right)^{-1},\\
          \widehat{\bmu}_{\bbeta} &\longleftarrow
          \left[\left(\frac{\widehat{a}_{\sigma_e^2}}{\widehat{b}_{\sigma_e^2}}\right)\bD_1(\mathbb{X})\bTheta + \left(\frac{\widehat{a}_{\sigma_e^2}}{\widehat{b}_{\sigma_e^2}}\right)\bD_2(\mathbb{X})(\eye_p - \bTheta) + \frac{\widehat{a}_{\sigma_0^2}}{\widehat{b}_{\sigma_0^2}}\eye_p\right]^{-1}\\
          &\quad\quad\times\left(
          \frac{\widehat{a}_{\sigma_e^2}}{\widehat{b}_{\sigma_e^2}}\bv + \frac{\widehat{a}_{\sigma_0^2}}{\widehat{b}_{\sigma_0^2}}\widehat{\mu}_{\mu_0}\one_p
          \right)
          .
        \end{align*}
        Note that the above vectorized updating rules can easily be implemented in \verb|R|, \verb|MATLAB|, or \verb|Python|. For example, in \texttt{R}, the updating formula for $\widehat{\bsigma}_{\bbeta}^2$ can be done using the command
\begin{verbatim}
sigma2_hat_beta[ , iter] = 
    (ahat_sigmae[iter - 1]/bhat_sigmae[iter - 1] * sum(X ^ 2) + 
     (ahat_sigma0[iter - 1]/bhat_sigma0[iter - 1])) ^ (-1)
\end{verbatim}
        \item Update $(\widehat{\bmu}_{\bb_i},\widehat{\bPsi}_{\bb_i})$ for all $i = 1,\ldots,m$:
        \begin{align*}
        \widehat{\bmu}_{\bb_i} &\longleftarrow \widehat{\bPsi}_{\bb_i}\left[\frac{\widehat{a}_{\sigma_e^2}}{\widehat{b}_{\sigma_e^2}}\sum_{j = 1}^{n_i}(\widehat{y}_{ij} - \bx_{ij}\transpose\bbeta)\bz_{ij}\right],
        \\
        \widehat{\bPsi}_{\bb_i} &\longleftarrow \left(\frac{\widehat{a}_{\sigma_e^2}}{\widehat{b}_{\sigma_e^2}}\sum_{j = 1}^{n_i}\bz_{ij}\bz_{ij}\transpose + \widehat\bPsi^{-1}\right)^{-1}.
        \end{align*}
        \item Update $(\widehat{a}_{\sigma_e^2}, \widehat{b}_{\sigma_e^2}), (\widehat{a}_{\sigma_0^2}, \widehat{b}_{\sigma_0^2}), (\widehat{a}_w,\widehat{b}_w), (\widehat{\mu}_{\mu_0}, \widehat{\sigma}_{\mu_0}^2)$:
        \begin{align*}
          &\text{SSR} \longleftarrow \sum_{i = 1}^m\sum_{j = 1}^{n_i}\left\{\langle{y}_{ij}^2\rangle_q +  \bz_{ij}\transpose(\widehat{\bmu}_{\bb_i}\widehat{\bmu}_{\bb_i}\transpose + \widehat{\bPsi}_{\bb_i})\bz_{ij} + 
          \left(\sum_{k = 1}^px_{ijk}\theta_k\widehat{\mu}_{\beta_k}\right)^2
          \right\}\\
          &\quad\quad\quad\quad + \sum_{i = 1}^m\sum_{j = 1}^{n_i}\left\{
          \sum_{k = 1}^px_{ijk}^2[\theta_k(1 - \theta_k)\widehat{\mu}_{\beta_k}^2 + \theta_k\widehat{\sigma}_{\beta_k}^2]
          \right\}\\
          &\quad\quad\quad\quad - 2\sum_{i = 1}^m\sum_{j = 1}^{n_i}\widehat{y}_{ij}\left(\sum_{k = 1}^p x_{ijk}\theta_k\widehat{\mu}_{\beta_k} + \bz_{ij}\transpose\widehat{\bmu}_{\bb_i}\right)\\
          &\quad\quad\quad\quad + 2\sum_{i = 1}^m\sum_{j = 1}^{n_i}\widehat{\bmu}_{\bb_i}\transpose\bz_{ij}\left(\sum_{k = 1}^p x_{ijk}\theta_k\widehat{\mu}_{\beta_k}\right),\\
          &\widehat{a}_{\sigma_e^2} \longleftarrow a_1 + \frac{1}{2}\sum_{i = 1}^mn_i,\quad
          \widehat{b}_{\sigma_e^2} \longleftarrow b_1 +  \frac{1}{2}\text{SSR},\\
          &\widehat{a}_{\sigma_0^2} \longleftarrow 1 + \frac{1}{2}\sum_{k = 1}^p\theta_k,
          \quad\widehat{b}_{\sigma_0^2} \longleftarrow 1 + \frac{1}{2}\sum_{k = 1}^p\theta_k[(\widehat{\mu}_{\mu_0} - \widehat{\mu}_{\beta_k})^2 + \widehat{\sigma}_{\mu_0}^2 + \widehat{\sigma}_{\beta_k}^2],\\
          &\widehat{a}_w = a_w + \sum_{k = 1}^p\theta_k,\quad \widehat{b}_w = b_w + \sum_{k = 1}^p(1 - \theta_k)\\
          &\widehat{\sigma}_{\mu_0}^2 \longleftarrow \left(1 + \frac{\widehat{a}_{\sigma_0^2}}{\widehat{b}_{\sigma_0^2}}\sum_{k = 1}^p\theta_k\right)^{-1},\quad \widehat{\mu}_{\mu_0} \longleftarrow \widehat{\sigma}_{\mu_0}^2\frac{\widehat{a}_{\sigma_0^2}}{\widehat{b}_{\sigma_0^2}}\sum_{k = 1}^p\theta_k\widehat{\mu}_{\beta_k}.
        \end{align*}
      \end{itemize}
      \item \textbf{Step 3:} Output variational parameters
      \begin{align*}
      \mathrel{\Big\{}&
(\widehat{\mu}_{y_{ij}}, \widehat{\sigma}_{y_{ij}}^2)_{i,j\in\calI_{(\mathrm{mis})}},(\theta_k, \widehat{\mu}_{\beta_k}, \widehat{\sigma}_{\beta_k}^2)_{k = 1}^p, (\widehat{\bmu}_{\bb_i}, \widehat{\bPsi}_{\bb_i})_{i = 1}^m, \\
    &(\widehat{a}_{\sigma_e^2}, \widehat{b}_{\sigma_e^2}), (\widehat{a}_{\sigma_e^2}, \widehat{b}_{\sigma_e^2}), 
    \widehat{\bPsi}, (\widehat{\mu}_{\mu_0}, \widehat{\sigma}_{\mu_0}^2),  (\widehat{a}_{w}, \widehat{b}_{w})
\mathrel{\Big\}}.
      \end{align*}
\end{enumerate}

\section{Application in Multiple Imputation of Missing Data} 
\label{sec:application_in_multiple_imputation_of_missing_data}

In this section, we discuss the application of the variational Bayesian inference algorithm developed in Section \ref{sub:variational_inference_with_spike_and_slab_} to multiple imputations of missing data. Note that in Section \ref{sec:linear_mixed_effects_regression_models}, the response variable is assumed to be continuous, and the sampling model of interest posits a linear mixed-effect model with Gaussian noise. We also consider a categorical response variable and apply a mis-specified linear mixed-effect model as an approximation to the posterior predictive distribution using variational Bayesian inference for the purpose of MI. To correct this misspecification, we apply a calibration-based method \cite{doi:10.1198/000313008X300912,doi:10.1002/sim.4355} using a set of rounding rules. 


Consider a collection data matrices $(\bM_i)_{i = 1}^m$, where $\bM_i$ is an $n_i\times d$ data matrix with rows representing observations and columns representing variables, and $i\in\{1,\ldots,m\}$ is the index for clusters. 
 We let $(\bM_i)_{(\mathrm{mis})}$ and $(\bM_i)_{(\mathrm{obs})}$ denote the missing and observed portion of $\bM_i$, respectively, and develop a multiple imputation method by leveraging the variational Bayesian inference method discussed in Section \ref{sub:variational_inference_with_spike_and_slab_}.
In an MCMC sampler, posterior samples of the model parameters $\bTheta$ are drawn from a Markov chain that converges in distribution to the posterior distribution of $\bTheta$ given $\bY_{(\mathrm{obs})}$. As such, posterior predictive samples of the $\bY_{(\mathrm{mis})}$, upon convergence, can serve as the imputed values of the missing portion of $\bY$. 
In what follows, we outline a strategy imputing $\bY_{(\mathrm{mis})}$ using the output of the variational inference algorithm in Section \ref{sub:variational_inference_with_spike_and_slab_}. 

We follow a sequential hierarchical regression imputation (SHRIMP) strategy developed in \cite{yucel2017sequential}, which has also been applied in \cite{QiushuangLi2020}. Following these implementations,  variables are assumed to be sorted by columns according to missingness rates:
for a variable index $k\in \{1,\ldots,d\}$, we define the missing ratio as the percentage of missingness:
\[
R_k^{(\mathrm{mis})}:= \frac{1}{N}\sum_{i = 1}^m\sum_{j = 1}^{n_i}\mathbbm{1}(M_{ijk}\text{ is NA}),
\]
where $M_{ijk}$ denotes the $(j, k)$th entry of $\bM_i$. 
Then SHRIMP relabels the column indices $[d] = \{1,2,\ldots,d\}$ as $\{k_1,\ldots,k_d\}$ such that
$R_{k_r}^{(\mathrm{mis})}\leq R_{k_{r + 1}}^{(\mathrm{mis})}$ for all $r = 1,2,\ldots,d - 1$. Namely, the ratio of the missing percentage of a column in $\bM$ is always no greater than that of the next column after sorting. Then, for each fixed $k$, SHRIMP sequentially imputes missing values of $\bM$ from $k = k_1$ to $k = k_n$ using the algorithm detailed below, which utilizes the output of the variational inference method. 

As linear mixed-effect model \eqref{eqn:LME_model} is designed for continuous responses, we ``calibrate" the imputations to better reflect the observed distributions for the caregorical data \citep{doi:10.1198/000313008X300912,doi:10.1002/sim.4355}. Their method allows this model misspecification while taking advantage of the computational convenience and is quite general imputation method for missing categorical data. The basic requirement is that there exists a surrogate imputation model $p(\bY_C)$ that generates continuous data such that the categorical data $\bY$ can be viewed as a discretized version of the continuous data $\bY_C$ generated from the surrogate imputation model $p(\bY_C)$. 

The key idea of the calibration-based imputation method can be loosely stated as follows. We generate two copies $\bY_C, \bY_{(\mathrm{dup}),C}$, using an imputation method based on the continuous surrogate model $p(\bY_C)$. Let $\bY_{(\mathrm{obs}),C}$, $\bY_{(\mathrm{mis}),C}$ be the portion of $\bY_C$ corresponding to the observed or missing locations of $\bY_{(\mathrm{obs})}$ or $\bY_{(\mathrm{mis})}$ in $\bY$, respectively, and we define $\bY_{(\mathrm{obs,dup}),C}$ and $\bY_{(\mathrm{mis,dup}),C}$ for $\bY_{(\mathrm{dup}),C}$ similarly. Suppose the number of total categories in $\bY$ is $G$, and the categories are labeled as $g = 1,2,\ldots,G$. Then, the calibration-based imputation method proposes to compute a sequence of cut-off values $-\infty = c_0, c_1,\ldots,c_{g - 1}$ that can be determined as follows. For each $(i, j)$ pair corresponding to an observed $y_{ij}$, we let $y_{ij}^{(\mathrm{dup})} = g$ if the $(i, j)$th element of $\bY_{(\mathrm{obs,dup}),C}$ lies in the interval $(c_{g - 1}, c_g]$. The cut-off values $c_1,\ldots,c_{g - 1}$ are selected such that for each category $g = 1,\ldots,G$, 
\[
\sum_{(i, j)\in \calI_{(\mathrm{obs})}}\mathbbm{1}\{y_{ij}^{(\mathrm{dup})} = g\} = \sum_{(i, j)\in \calI_{(\mathrm{obs})}}\mathbbm{1}\{y_{ij} = g\}. 
\]
In other words, the proportions of different categories after applying the cut-off values $c_0,c_1,\ldots,c_{g - 1}$ to $\bY_{(\mathrm{obs,dup}),C}$ matches with the proportions of different categories in $\bY_{(\mathrm{obs})}$. These cut-off values can be computed explicitly using quantiles of $\bY_{(\mathrm{obs,dup}),C}$. 

We now incorporate the above calibration-based rounding strategy with the sequential VBMI method discussed in Section \ref{sec:application_in_multiple_imputation_of_missing_data} through the following algorithm. Note that the algorithm below can deal with the case where the data contains both continuous variables and categorical variables (see Step 4 for details). 
Note that the algorithm below only generate one imputation copy, and one needs to repeat it $M$ independent times to obtain $M$ copies of imputed versions of $(\bM_i)_{i = 1}^m$. 
\begin{enumerate}
\item[$\blacksquare$] \textbf{Input: }
            
\item Sort the variable indices $\{1,\ldots,d\}$ as $\{k_1,\ldots,k_d\}$ such that $R_{k_r}^{(\mathrm{mis})}\leq R_{k_{r + 1}}^{(\mathrm{mis})}$ for all $r = 1,\ldots, d - 1$. 

\item Use some imputation method to fill the missing values in $\bM_1,\ldots,\bM_m$ (e.g., the predictive mean matching). 

\item \textbf{For }$k = k_1,\ldots,k_d$:

Set $y_{ij} = M_{ijk}$, $\bx_{ij} = (M_{ijl}:l\in\{1,\ldots,d\},l\neq k)$, and $\bz_{ij}$ the $i$th standard basis vector in $\mathbb{R}^m$ with the $i$th entry being $1$ and the remaining entries being $0$. Call the variational Bayesian inference algorithm in Section \ref{sub:variational_inference_with_spike_and_slab_} to obtain the variational parameters
      \begin{align*}
      \mathrel{\Big\{}&
(\widehat{\mu}_{y_{ij}}, \widehat{\sigma}_{y_{ij}}^2)_{i,j\in\calI_{(\mathrm{mis})}},(\theta_k, \widehat{\mu}_{\beta_k}, \widehat{\sigma}_{\beta_k}^2)_{k = 1}^p, (\widehat{\bmu}_{\bb_i}, \widehat{\bPsi}_{\bb_i})_{i = 1}^m, \\
    &(\widehat{a}_{\sigma_e^2}, \widehat{b}_{\sigma_e^2}), (\widehat{a}_{\sigma_e^2}, \widehat{b}_{\sigma_e^2}), 
    \widehat{\bPsi}, (\widehat{\mu}_{\mu_0}, \widehat{\sigma}_{\mu_0}^2),  (\widehat{a}_{w}, \widehat{b}_{w})
\mathrel{\Big\}}.
      \end{align*}
Sample
      \begin{align*}
      &\sigma_e^2\sim\mathrm{IG}(\widehat{a}_{\sigma_e^2}, \widehat{b}_{\sigma_e^2}),\\
      &\beta_k\sim \theta_k\mathrm{N}(\widehat{\mu}_{\beta_k}, \widehat{\sigma}_{\beta_k}^2) + (1 - \theta_k)\delta_0,\quad k = 1,\ldots,p,\\
&\bb_i\sim \mathrm{N}(\widehat{\bmu}_{\bb_i}, \widehat{\bPsi}_{\bb_i}),\quad i = 1,\ldots,m,\\
&\widetilde{y}_{ij}\sim\mathrm{N}(\bx_{ij}\transpose\bbeta + \bz_{ij}\transpose \bb_i, \sigma_{e}^2)\quad\mbox{if }(i, j)\in\mathcal{I}_{(\mathrm{mis})}
        \end{align*}
If the $k$th variable is a continuous variable, then update $M_{ijk} = \widetilde{y}_{ij}$ if $(i, j)\in\mathcal{I}_{(\mathrm{mis})}$. 

If the $k$th variable is a categorical variable with $G$ categories, then sample
\begin{align*}
      y^{(\mathrm{dup}),C}_{ij} &\sim\mathrm{N}(\bx_{ij}\transpose\bbeta + \bz_{ij}\transpose \bb_i, {\sigma}_{e}^2),\quad
      y^{C}_{ij} \sim\mathrm{N}(\bx_{ij}\transpose\bbeta + \bz_{ij}\transpose \bb_i, {\sigma}_{e}^2),
      \end{align*}
for all $i\in\{1,\ldots,m\}$ and $j\in\{1,\ldots,n_i\}$. For each $g = 1,2,\ldots,G - 1$, compute 
      \[
      s_g = \frac{1}{|\calI_{(\mathrm{obs})}|}\sum_{(i,j)\in\calI_{(\mathrm{obs})}}\mathbbm{1}\{y_{ij}\leq g\}.
      \]
      Compute the $s_g$-sample quantile $c_g$ of $\{y^{\mathrm{(dup)}, C}_{ij}:(i, j)\notin\calI_{(\mathrm{mis})}\}$ such that
      \[
      s_g = \frac{1}{ |\calI_{(\mathrm{obs})}|}\sum_{(i,j)\in\calI_{(\mathrm{obs})}}\mathbbm{1}\{y^{(\mathrm{dup}),C}_{ij} \leq c_g\},\quad g = 1,\ldots,G - 1.
      \]
      Set $c_0 = -\infty$ and $c_G = +\infty$ and update $M_{ijk} = g$ if $(i, j)\in\mathcal{I}_{(\mathrm{mis})}$ and $y_{ij}^C\in(c_{g - 1}, c_g]$.
\item \textbf{Output:} $(\bM_i)_{i = 1}^m$.
\end{enumerate}



\section{Numerical examples} 
\label{sec:numerical_examples}

\subsection{Continuous data example} 
\label{sub:continuous_data_example}

\begin{figure}[htbp]
  \centerline{\includegraphics[width=0.7\textwidth]{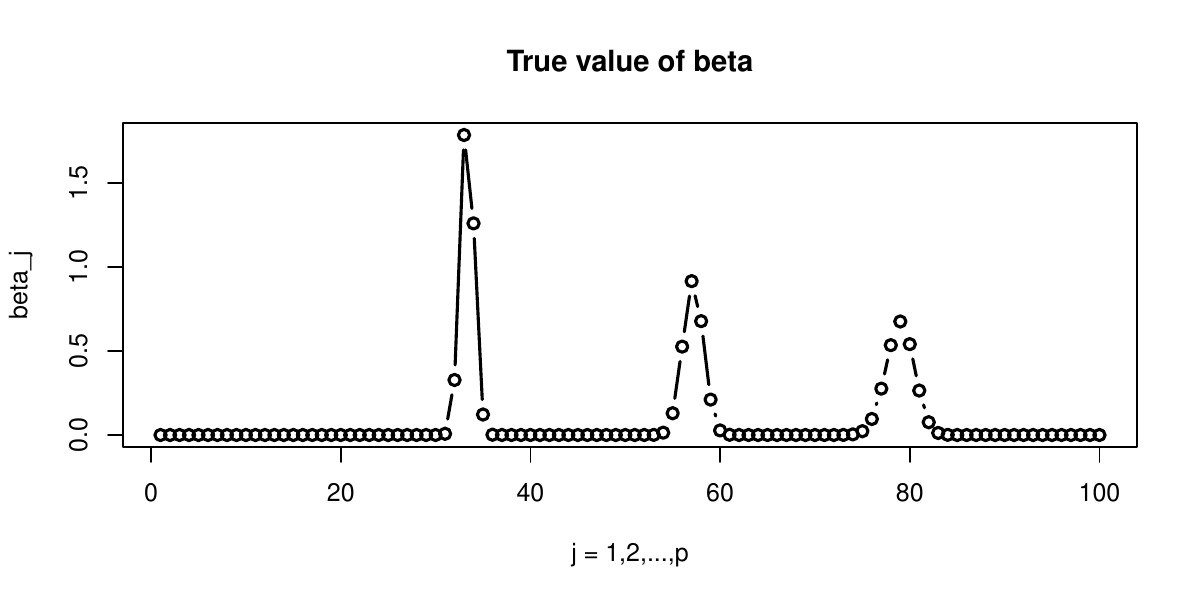}}
  \caption{True value of $\bbeta = [\beta_1,\ldots,\beta_p]$ in Section \ref{sub:continuous_data_example}. }
  \label{fig:Continuous_beta_TV}
\end{figure}

We assume that the data generation mechanism for the synthetic data is the same as \eqref{eqn:LME_model}:
\[
y_{ij} = \bx_{ij}\transpose{}\bbeta + \bz_{ij}\transpose{}\bb_i + \eps_{ij},\quad \eps_{ij}\iidsim\mathrm{N}(0, \sigma_e^2), \quad \bb_i\iidsim\mathrm{N}(\zero_l, \bPsi),
\]
where $i = 1,\ldots,m$, $j = 1,\ldots,n_i$. We set $m = 50$, $n_i = n = 20$, $p = 100$, and $l = 3$. The entries of the fixed-effect covariates $\bx_{ij}$'s are independently generated from $\mathrm{N}(0, 3^2)$ for all $i, j$, and the entries of the random-effect covariates $\bz_{ij}$'s are independently drawn from $\mathrm{N}(0, 1)$ for all $i, j$. The fixed-effect regression coefficient $\bbeta$ is assumed to have a weakly sparse structure. We adopt the ``three-peak curve'' example constructed in \cite{johnstone2009consistency} and set the entries of $\bbeta = [\beta_1,\ldots,\beta_p]\transpose{}\in\mathbb{R}^p$ as follows:
\begin{align*}
\beta_k = &\frac{0.7}{20}\mathrm{Beta}\left( k/p \mid 1500, 3000\right) + \frac{0.5}{20}\mathrm{Beta}(k/p \mid 1200, 900)\\
&\quad + \frac{0.5}{20}\mathrm{Beta}(k/p \mid 600, 160),
\end{align*}
where $\mathrm{Beta}(t\mid a, b)$ is the density of the $\mathrm{Beta}(a, b)$ distribution evaluated at $t\in [0, 1]$. Figure \ref{fig:Continuous_beta_TV} below visualizes the true values of $\beta_j$ as a function of $j = 1,\ldots,p$, from which we can see that a significant portion of the entries of $\bbeta$ are rather close to $0$, whereas only several entries are bounded away from $0$. The covariance matrix $\bPsi$ for the random effect is set to be the $3\times 3$ identity matrix for simplicity.

We next impose missing values on the response data $\by = (y_{ij}:i\in\{1,\ldots,m\},j\in\{1,\ldots,n_i\})$ under missing at random mechanism, meaning that the distribution of the missingness depends on the observed variables but not the missing values. To this end, we generate $\bR_\by = (R_{y_{ij}}:i\in\{1,\ldots,m\},j\in\{1,\ldots,n\})$ to assign missing values to $\by$. Specifically, we take
\begin{align*}
R_{y_{ij}}\sim\mathrm{Bernoulli}(p_{ij}),\quad \mathrm{logit}(p_{ij}) = \alpha_R + \beta_{\mathrm{mis}}x_{ij1}.
\end{align*}
Here, the parameters $\alpha_R$ and $\beta_{\mathrm{mis}}$ are tunned such that the overall missing percentage of $\by$ is approximately $27\%$, i.e., $1/(mn)\sum_{i = 1}^m\sum_{j = 1}^nR_{y_{ij}}\approx 0.27$. We then set $y_{ij}$ to be \verb|NA| if $R_{y_{ij}} = 1$, and maintain the original value of $y_{ij}$ if $R_{y_{ij}} = 0$. 

\begin{figure}[htbp]
  \centerline{\includegraphics[width=1\textwidth]{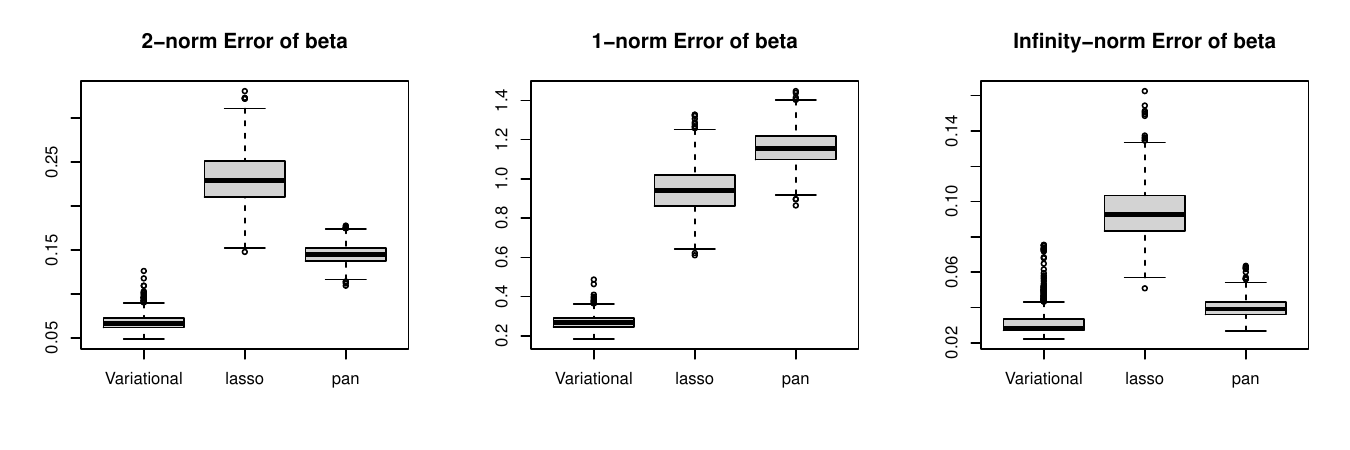}}
  \caption{The boxplots of the 2-norm errors, the 1-norm errors, and the infinity-norm errors for estimating $\bbeta$ using the variational Bayesian inference method, the lasso, and the \texttt{pan} package for the simulated example in Section \ref{sub:continuous_data_example}. }
  \label{fig:beta_boxplot_continuous}
\end{figure}

The inference tasks here are twofold: Parameter estimation for $\bbeta$ and multiple imputation for $\bY_{(\mathrm{mis})}$. We first focus on the performance of the parameter estimation for the fixed-effect regression coefficient $\bbeta$ in the presence of the potential missing values in $\by$. To accomplish this, we implemented the proposed variational Bayesian inference algorithm in Section \ref{sub:variational_inference_with_spike_and_slab_} and compare it with the classical lasso method \citep{https://doi.org/10.1111/j.2517-6161.1996.tb02080.x} implemented in the \verb|glmnet| package and the \verb|pan| package \citep{zhao2013pan}. The entire experiment was repeated for $1000$ Monte Carlo replicates. For each replicate of the synthetic dataset, we computed three types of estimation error for the estimate $\widehat{\bbeta}$ obtained from the three methods: The 2-norm error $\|\widehat{\bbeta} - \bbeta\|_2 = [\sum_k(\widehat{\beta}_k - \beta_k)^2]^{1/2}$, the 1-norm error $\|\widehat{\bbeta} - \bbeta\|_1 = \sum_k|\widehat{\beta}_k - \beta_k|$, and the infinity-norm error $\|\widehat{\bbeta} - \bbeta\|_\infty = \max_k|\widehat{\beta}_k - \beta_k|$. Here, $\widehat{\bbeta}$ represents a generic estimate for $\bbeta$. For the variational Bayesian inference method, $\widehat{\bbeta}$ was taken to be the variational posterior mean; For the pan package, we considered $\widehat{\bbeta}$ to be the posterior mean. Below, Figure \ref{fig:beta_boxplot_continuous} visualizes the boxplots of the three types of estimation errors across the $1000$ Monte Carlo replicates for the three methods. 
We can see that in terms of the 2-norm errors and the 1-norm errors, the variational posterior mean is significantly smaller than the other two competitors. In terms of the infinity-norm errors, the variational posterior mean is slightly better than the \verb|pan| package, and both are also significantly better than the lasso estimate. We also remark that both the variational Bayesian inference method and the \verb|pan| package provide natural environments for dealing with missing responses and can perform multiple imputations. This part of the analysis demonstrates the advantage of the proposed method in terms of parameter estimation in the presence of missing responses and sparsity. In addition, we also provide the visualization of the point estimates using the variational Bayesian inference method, the lasso, and the \verb|pan| package in a randomly selected replicate in Figure \ref{fig:Continuous_beta} below. From the perspective of a single synthetic dataset, the performance of the lasso is similar to the variational Bayesian inference method, but the \verb|pan| package provides a worse estimate because the sparsity structure of $\bbeta$ is not captured. 
\begin{figure}[htbp]
  \centerline{\includegraphics[width=1\textwidth]{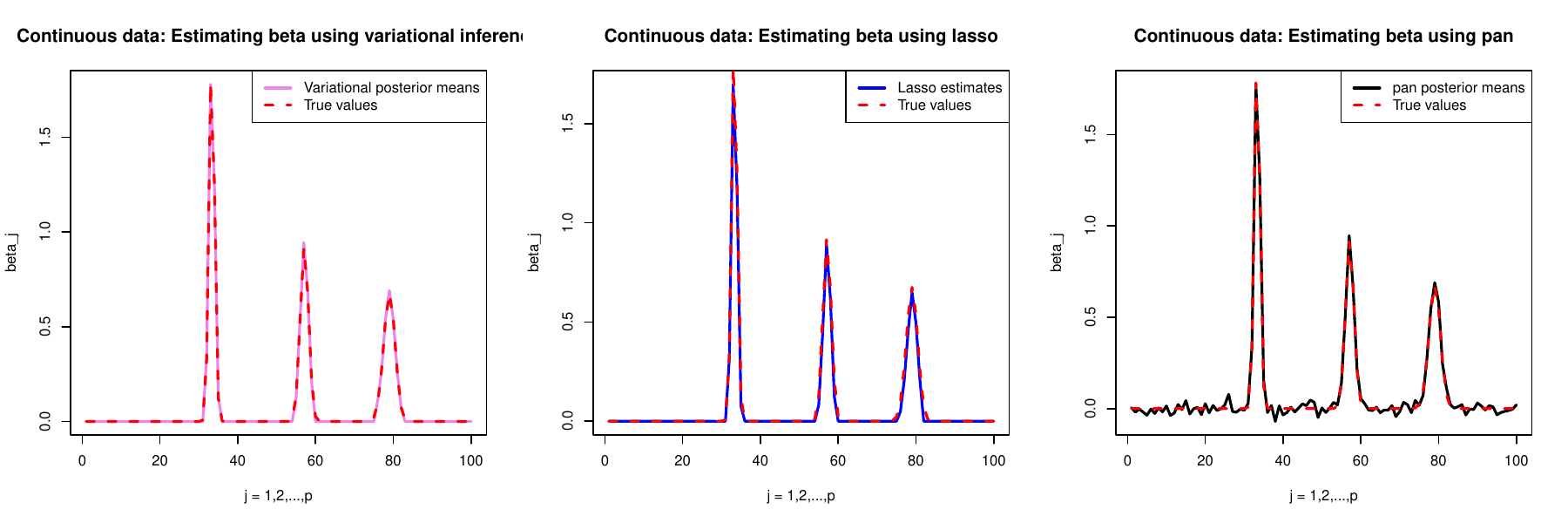}}
  \caption{Visualization of the estimation for $\bbeta$ using the variational Bayesian inference method, the lasso, and the \texttt{pan} package for the simulated example in Section \ref{sub:continuous_data_example}. }
  \label{fig:Continuous_beta}
\end{figure} 

In addition, we computed the $95\%$-confidence intervals for the entries of the regression coefficients $\bbeta$ whose absolute values are no less than $\tau = 0.05$ using Rubin's combined rule \citep{rubinmultiple}. The empirical coverage probabilities of these $95\%$-confidence intervals for the regression coefficients $\beta_j$'s with $|\beta_j| > 0.05$ are tabulated in Table \ref{tab:coverage_probability_beta_continuous} below. It can be clearly seen that the empirical coverage probabilities are all close to the nominal $95\%$ probability.

\begin{table}[htbp]
\centering
\caption{Coverage probability for the MI estimates for $\beta_j$ (with $|\beta_j| \geq 0.05$) using the variational Bayes multiple imputation (VBMI) method for the simulated example in Section \ref{sub:continuous_data_example}. Here, $\{j\in\{1,\ldots,100\}:|\beta_j|\geq 0.05\} = \{32,33,34,35,55,56,57,58,59,76,77,78,79,80,81,82\}$(reall $\bbeta$ is a $100$-dimensional vector but only a small fraction of $|\beta_j|$'s are larger than $0.05$). }
\begin{tabular}{c c c c c c c c c}
\hline\hline
$\beta_j$     & $\beta_{32}$ & $\beta_{33}$ & $\beta_{34}$ & $\beta_{35}$ & $\beta_{55}$ & $\beta_{56}$ & $\beta_{57}$ & $\beta_{58}$ \\
\hline
Coverage   & 95.9\% & 94.1\% & 94.8\% & 93.2\% & 94.2\% & 93.3\% & 94.3\% & 93.3\% 
\\
\hline
& $\beta_{59}$ & $\beta_{76}$ & $\beta_{77}$ & $\beta_{78}$ & $\beta_{79}$ & $\beta_{80}$ & $\beta_{81}$ & $\beta_{82}$
\\
\hline
Coverage & 95.5\% & 94.0\% & 94.5\% & 94.6\% & 94.3\% & 93.7\% & 94.6\% & 94.9\%\\
\hline
\hline
\end{tabular}
\label{tab:coverage_probability_beta_continuous}
\end{table}

\begin{figure}[htbp]
  \centerline{\includegraphics[width=0.9\textwidth]{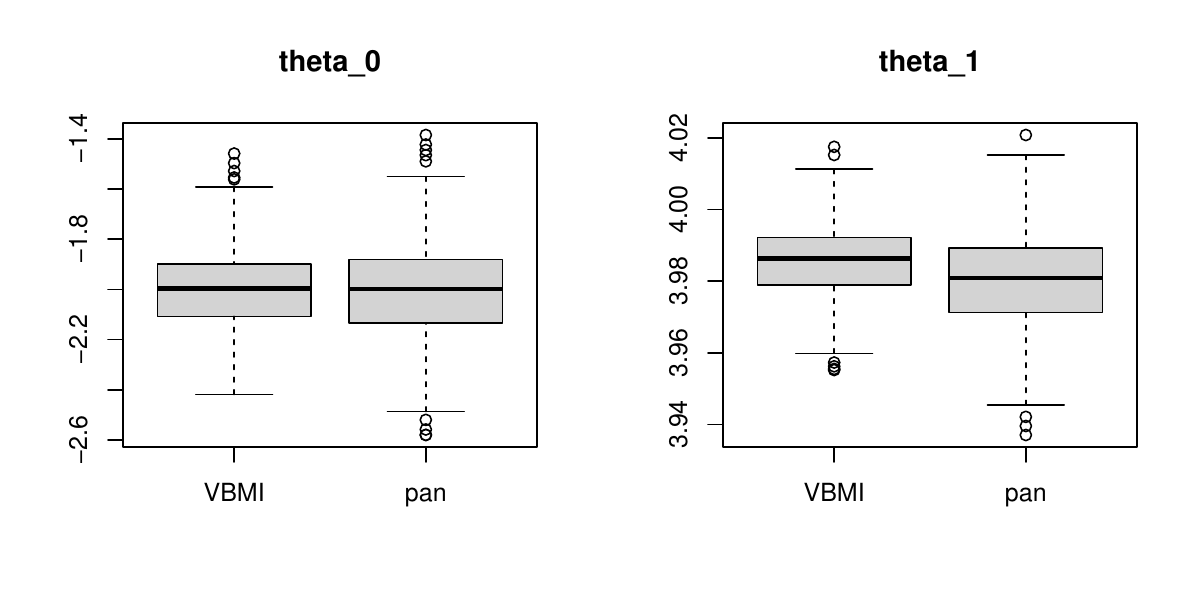}}
  \caption{Visualization of the boxplots of $\theta_0$ and $\theta_1$ using the variational Bayes multiple imputation (VBMI) method and the \texttt{pan} package for the simulated example in Section \ref{sub:continuous_data_example}. }
  \label{fig:theta_boxplot_continuous}
\end{figure}

\begin{figure}[htbp]
  \centerline{\includegraphics[width=0.9\textwidth]{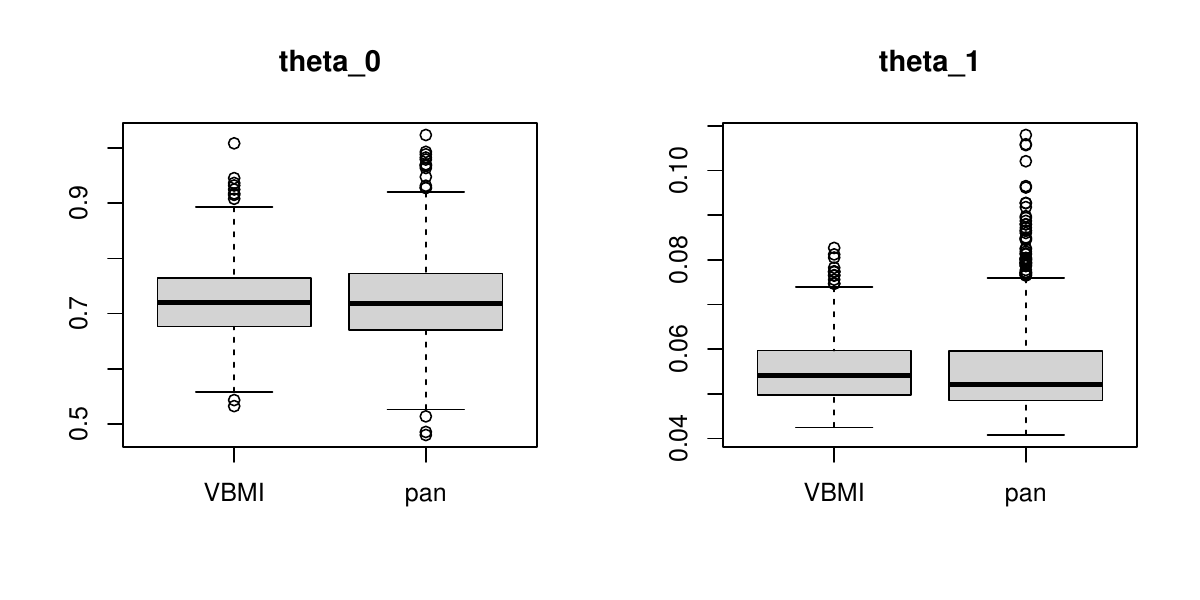}}
  \caption{Visualization of the boxplots of the confidence interval widths for $\theta_0$ and $\theta_1$ using the variational Bayes multiple imputation (VBMI) method and the \texttt{pan} package for the simulated example in Section \ref{sub:continuous_data_example}. }
  \label{fig:theta_CIwidth_boxplot_continuous}
\end{figure}

We next investigate the performance of the multiple imputations. In preparation for doing so, we develop another layer of the conditional model using the response data $\by$ as the covariate. Specifically, we consider the following (conditional) linear mixed-effect model:
\begin{align*}
u_{ij} = \theta_0 + \theta_1y_{ij} + v_i + e_{ij},\quad v_1,\ldots,v_m\iidsim\mathrm{N}(0, 1),\quad e_{ij}\iidsim\mathrm{N}(0, 0.3^2),
\end{align*}
$i = 1,\ldots,m$, $j = 1,\ldots,n$, where we set the fixed-effect coefficient as $\theta_0 = -2$ and $\theta_1 = 4$. Here, the matrix $\by$ contains missing values but the response variables $u_{ij}$ are generated using the complete data. The aforementioned $1000$ Monte Carlo replicates are applied here to generate $u_{ij}$'s, $i = 1,\ldots,m,j = 1,\ldots,n$. For the imputation methods, we implement the proposed VBMI strategy in Section \ref{sec:application_in_multiple_imputation_of_missing_data}, together with the \verb|pan| package, for comparison. The number of imputations is set to $M = 5$. As the focus here is to evaluate the performance of the multiple imputation inference, we choose to use the \verb|lm| function in \verb|R| to perform the regression analysis for $\theta_0$ and $\theta_1$ for simplicity. After obtaining the estimates for $\theta_0$ and $\theta_1$ using the \verb|lm| function in \verb|R|, we apply Rubin's rule for combined analysis \citep{rubinmultiple}. Figure \ref{fig:theta_boxplot_continuous} below demonstrates the boxplots of the point estimates for $\theta_0$ and $\theta_1$ based on the variational Bayes multiple imputation (VBMI) method developed in Section \ref{sec:application_in_multiple_imputation_of_missing_data}, and the \verb|pan| package imputation estimates. We can see that these two methods provide similar performance in terms of the point estimates. 
Furthermore, Figure \ref{fig:theta_CIwidth_boxplot_continuous} presents the widths of the confidence interval for $\theta_0$ and $\theta_1$ computed using Rubin's combined analysis across the $1000$ Monte Carlo replicates. The performance here is also similar, with the \verb|pan| package producing slightly wider confidence intervals. 
However, the key difference between the two methods is in the coverage probabilities of the confidence intervals, which are tabulated in Table \ref{tab:coverage_probability_theta_continuous} below. The coverage probabilities are comparatively smaller than the nominal coverage probability $95\%$ due to the model misspecification. Still, the VBMI method outperforms the \texttt{pan} package with higher coverage probabilities for both $\theta_0$ and $\theta_1$, demonstrating the power of the proposed method. 
\begin{table}[htbp]
\centering
\caption{Coverage probability of the confidence intervals for $\theta_0$ and $\theta_1$ using the variational Bayes multiple imputation (VBMI) method and the \texttt{pan} package for the simulated example in Section \ref{sub:continuous_data_example}.}
\begin{tabular}{c c c }
\hline\hline
$\theta_j$     & VBMI coverage probability & \texttt{pan} coverage probability\\
\hline
$\theta_0$    & \textbf{86.6\%}  &  82.9\% \\
$\theta_1$    & \textbf{89.6\%} & 75.3\% \\
\hline\hline
\end{tabular}
\label{tab:coverage_probability_theta_continuous}
\end{table}

\subsection{Categorical data example} 
\label{sub:categorical_data_example}

We next consider examples with categorical incomplete data. Unlike the setup in Section \ref{sub:continuous_data_example}, the imputation models model \eqref{eqn:LME_model} is purposefully misspecified.
We consider the following two scenarios for the true data generation mechanism:
\begin{itemize}
  \item \textbf{Binary data} The generative model of the synthetic data is a logistic mixed-effect model:
  \[
  y_{ij} \sim\mathrm{Bernoulli}(p_{ij}),\quad
  \mathrm{logit}(y_{ij}) = \bx_{ij}\transpose{}\bbeta + \bz_{ij}\transpose \bb_i,\quad \bb_1,\ldots,\bb_m\iidsim\mathrm{N}_l(\zero_l, \bPsi),
  \]
where $i = 1,\ldots,m$, $j = 1,\ldots,n$, $m = 50$, $n_i = n = 20$, $p = 100$, $l = 3$, and $\bPsi = \eye_l$. We follow the setup in Section \ref{sub:continuous_data_example} and generate the entries of $\bx_{ij}$ from $\mathrm{N}(0, 3^2)$ and the entries of $\bz_{ij}$ from $\mathrm{N}(0, 1)$, independently. The fixed-effect regression coefficient $\bbeta$ is generated as follows: We first generate the entries of $\balpha = [\alpha_1,\ldots,\alpha_{10}]\transpose$ independently from $\mathrm{N}(0, 0.1^2)$, and then set $\beta_k = \alpha_k/\|\balpha\|_2$ with $k = 1,\ldots,10$, and $\beta_k = 0$ for $k = 11,\ldots,p = 100$. 

  \item \textbf{Setup 2: Categorical data with $5$ categories.} The generative model of the synthetic data is a multinomial mixed-effect model:
  \[
  \prob(y_{ij} = s\mid\bbeta_1,\ldots,\bbeta_K, \bB) = \frac{\exp(\bx_{ij}\transpose{}\bbeta_s + \bz_{ij}\transpose{}\bb_i)}{\sum_{t = 1}^K\exp(\bx_{ij}\transpose{}\bbeta_t + \bz_{ij}\transpose{}\bb_i)},
  \]
  where $i = 1,\ldots,m$, $j = 1,\ldots,n$, $K = 5$ is the number of categories, $m = 50$, $n_i = n = 20$, $p = 100$, $l = 3$, and $\bPsi = \eye_l$. The setup here is similar to Setup 1 above: The entries of $\bx_{ij}$ are generated from $\mathrm{N}(0, 3^2)$, and the entries of $\bz_{ij}$ are simulated from $\mathrm{N}(0, 1)$, independently. The fixed-effect regression coefficients $\bbeta_s = [\beta_{s1},\ldots,\beta_{sp}]\transpose{}$ for $s = 1,\ldots,K$ are set similar to the $\bbeta$ in Setup 1 above: We first draw the entries of $\balpha_s = [\alpha_{s1},\ldots,\alpha_{s,10}]\transpose$ independently from $\mathrm{N}(0, 0.1^2)$, and then set $\beta_{sk} = \alpha_{sk}/\|\balpha_s\|_2$ with $k = 1,\ldots,10$, and $\beta_{sk} = 0$ for $k = 11,\ldots,p = 100$. The generative process is repeated independently for each $s = 1,2,3,4,5$. 
\end{itemize}
In both scenarios, we impose missing values in $\bY$ under a missing at random (MAR) mechanism similar to Section \ref{sub:continuous_data_example}. We the set the overall missing percentage of the response data $\by$ is approximately $27\%$. The entire experiment for each setup above was repeated for $1000$ Monte Carlo replicates. 

As the generative model of the synthetic data differs from the imputation model \eqref{eqn:LME_model} employed here (i.e., model misspecification), we turn our focus to the performance of the multiple imputation inference rather than the parameter estimation. To this end, we follow the idea in Section \ref{sub:continuous_data_example} and use the response data to form an $m\times n$ matrix $\bY = [y_{ij}]_{m\times n}$ as the covariate matrix to construct a second-layer linear model. Specifically, given $\bY$, we consider
\begin{align}\label{eqn:second_layer_model_categorical}
\bu = \bY\btheta + \mathbf{e},\quad \mathbf{e}\sim\mathrm{N}_m(\zero_m, 5^2\eye_m).
\end{align}
Here, the second layer response variable $\bu = [u_1,\ldots,u_m]\transpose$ is generated from the complete data $\bY$, but the observable $\bY$ contains missing values. The regression coefficient $\btheta$ is generated from $\mathrm{N}_n(\zero_n, 5^2\eye_n)$. We repeat this $1000$ times. 
For comparison, we consider the \verb|pan| package and the rounding to the nearest integer strategy based on \verb|pan| package. The number of imputation datasets is set as $M = 5$, followed by Rubin's combined analysis for inference on $\btheta$. The inference for $\btheta$ is based on the \texttt{lm} function in \texttt{R} for standard linear models.

\begin{figure}[htbp]
  \centerline{\includegraphics[width= 1 \textwidth]{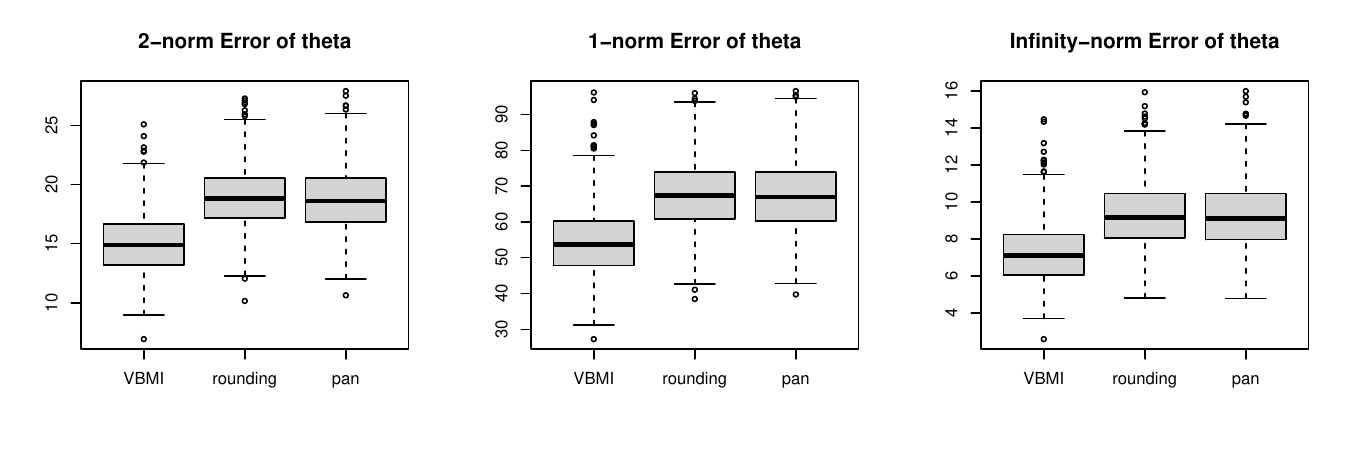}}
  \caption{The boxplots of the 2-norm errors, the 1-norm errors, and the infinity-norm errors for estimating $\btheta$ using the VBMI method, the rounding strategy, and the \texttt{pan} package for the simulated example (setup 1) in Section \ref{sub:categorical_data_example}. }
  \label{fig:theta_boxplot_binary}
\end{figure}

\begin{figure}[htbp]
  \centerline{\includegraphics[width= 1 \textwidth]{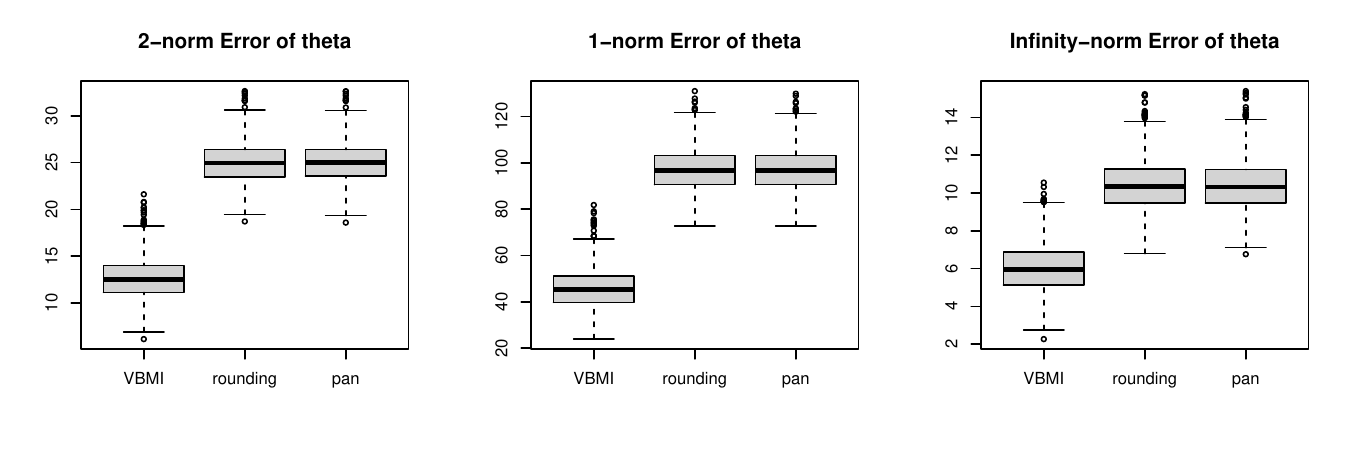}}
  \caption{The boxplots of the 2-norm errors, the 1-norm errors, and the infinity-norm errors for estimating $\btheta$ using the VBMI method, the rounding strategy, and the \texttt{pan} package for the simulated example (setup 2) in Section \ref{sub:categorical_data_example}. }
  \label{fig:theta_boxplot_categorical}
\end{figure}

\begin{figure}[htbp]
  \centerline{\includegraphics[width=0.7\textwidth]{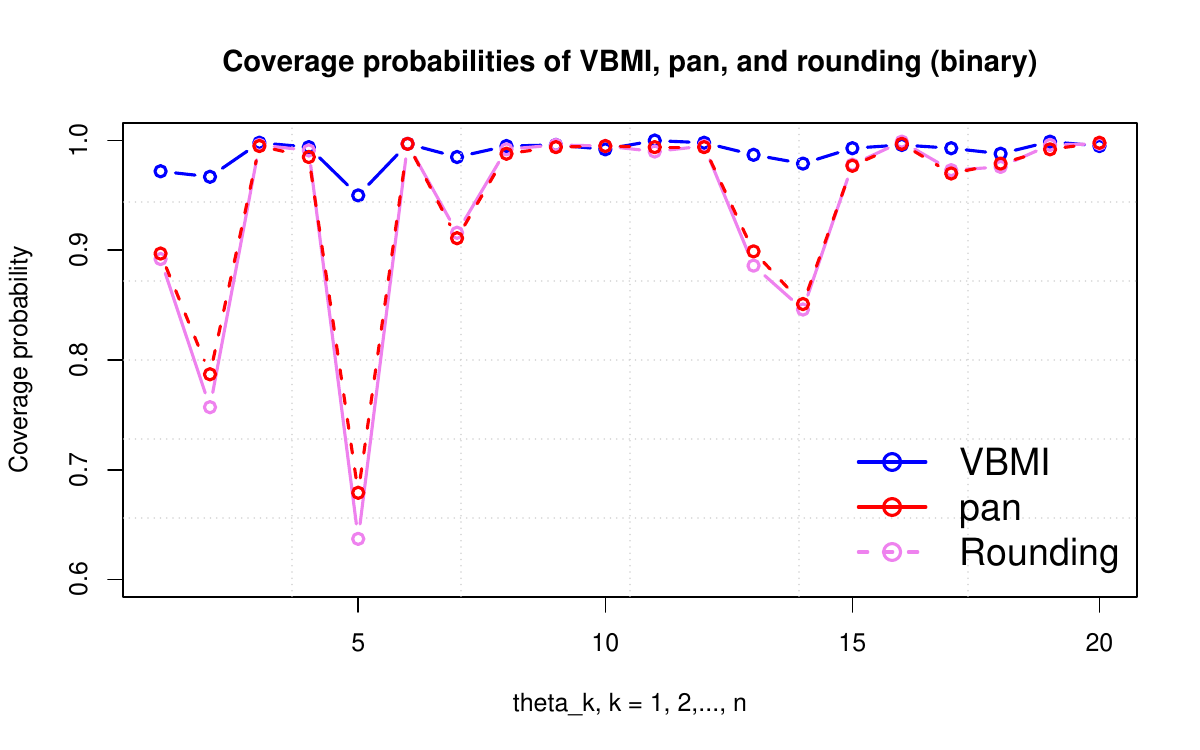}}
  \caption{Coverage probabilities of the confidence intervals for $\btheta$ using the variational Bayes multiple imputation (VBMI) method, the \texttt{pan} package, and the rounding strategy, for the simulated example (setup 1) in Section \ref{sub:categorical_data_example}. }
  \label{fig:CR_binary}
\end{figure}

\begin{figure}[htbp]
  \centerline{\includegraphics[width=0.7\textwidth]{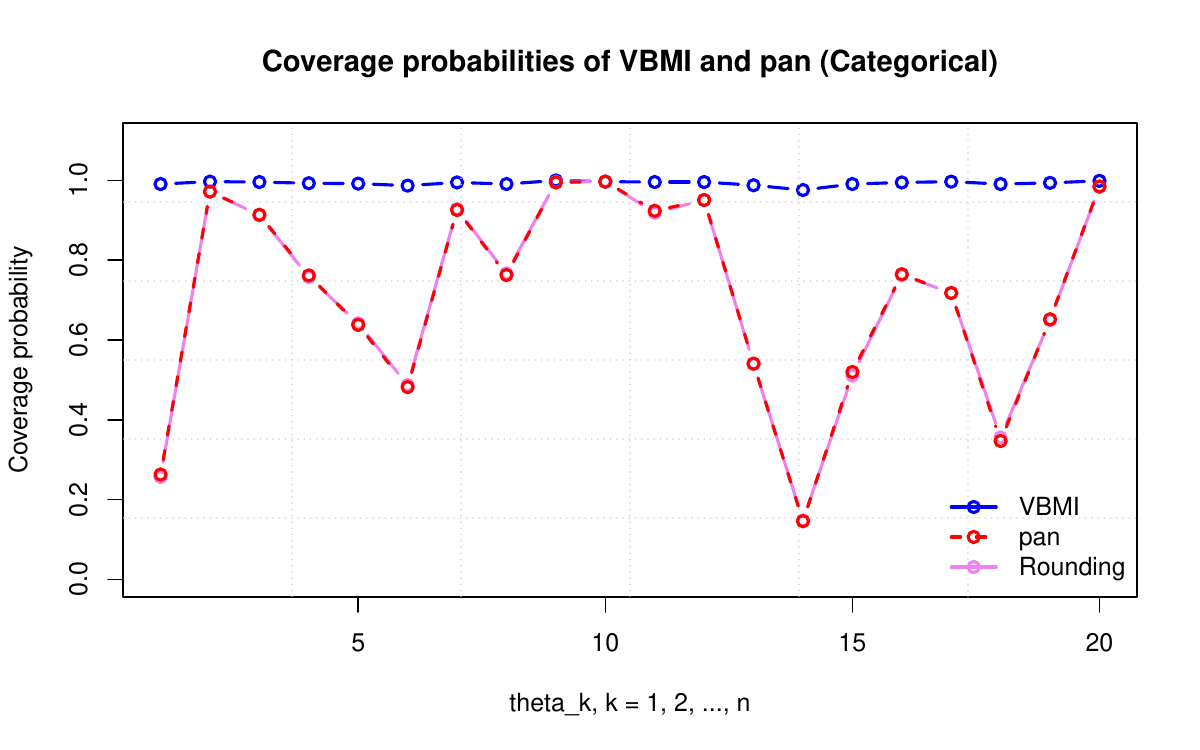}}
  \caption{Coverage probabilities of the confidence intervals for $\btheta$ using the variational Bayes multiple imputation (VBMI) method, the \texttt{pan} package, and the rounding strategy for the simulated example (setup 2) in Section \ref{sub:categorical_data_example}. }
  \label{fig:CR_categorical}
\end{figure}

Below, Figure \ref{fig:theta_boxplot_binary} and Figure \ref{fig:theta_boxplot_categorical} present the boxplots of the 2-norm errors, the 1-norm errors, and the infinity-norm errors for estimating $\btheta$ using different imputation methods under setup 1 and setup 2, respectively. 
It is clear that estimation error using the VBMI method is significantly smaller than the other two competitors in both setups for all three types of errors. 
Furthermore, Figure \ref{fig:CR_binary} and Figure \ref{fig:CR_categorical} visualize the empirical coverage probabilities of the confidence intervals across the $1000$ Monte Carlo replicates using the three imputation methods for setup 1 and setup 2, respectively. 
We can see that the coverage probabilities of the confidence intervals using the proposed VBMI method are consistent and satisfactory in comparison with the other two competitors. In particular, the \verb|pan| package rounding strategy is rather inconsistent and produces confidence intervals that are not reliable because the coverage probabilities are significantly lower than the nominal coverage $95\%$. Therefore, through empirical demonstration via the analyses of synthetic datasets, we show that the proposed VBMI method is powerful and robust in terms of the imputation for categorical responses even when the model is misspecified.



\subsection{Application to the National Survey of Children's Health Data}

Now we present the application of the sequential variational Bayesian multiple imputation (VBMI) strategy discussed in Section \ref{sec:application_in_multiple_imputation_of_missing_data} to a real-world National Survey of Children's Health (NSCH) data, which is available at \url{https://www.cdc.gov/nchs/slaits/cshcn.htm}. The data consists of the national survey data of children with special health care needs, and it includes a broad range of relevant demographical data, such as medical homes, health insurance, ages, races, sex, and education level, among others. We focus on the fiscal year 2020, where the dataset consists of the survey data of children with special care needs from 50 states and Washington D.C., and each cluster is defined to be a state or Washington D.C. with $m = 51$ clusters and approximately $750$ observations per cluster. 
Similar to many other survey data,
the NSCH data also have incomplete data, and drawing inference using the NSCH dataset with the VBMI algorithm is the focus of this subsection. Specifically, we focus on $37$ variables that are closely related to the children's special health needs and run the sequential VBMI algorithm discussed in Section \ref{sec:application_in_multiple_imputation_of_missing_data} with $M = 10$ multiple imputation copies. Each imputation copy can be regarded as a ``completed'' version of the dataset, and we draw further inferences for each variable based on these imputation copies separately and aggregate the inferential results finally through Rubin's combining rules \cite{rubinmultiple}. For comparison, we also consider the following ad-hoc analysis: We first collect observations with observed ``children with special health need (CSHCN)'' indicators and then compute the empirical probabilities of the categorical variables that are considered closely related to CSHCN after removing the missing values, together with the corresponding confidence intervals. These variables include the CSHCN, children's ages, races, sex, adults' education levels, and insurance types. The above results are tabulated in Table \ref{tab:NSCH_VBMI}, from which we can see that the VBMI creates inference that corrects the bias caused by the ad-hoc estimates for these empirical probabilities.

\begin{table}[htbp]
	\centering
	\caption{Comparison between the MI analysis and the ad-hoc analysis for the population means of selected variables (in percentages) }
\begin{tabular}{l|cc|ccc}
		\hline\hline
        &\multicolumn{2}{c|}{Ad-hoc analysis} & 
        \multicolumn{3}{c}{MI analysis}\\
	    &Mean&95\% CI&Mean&95\% CI&FMI ($\%$)\\
		\hline
		Special Health Need & & & & & \\ 
        (missing rate: 0\%) &23.71&(23.26, 24.16)& 23.48 & (23.08, 23.88) & NA\\
		\hline
        \multicolumn{6}{c}{Children Age (missing rate: 0\%)} \\
        \hline
        0-5 &29.04&(28.57, 29.53)& 28.38 & (27.95, 28.81) & NA \\
        6-12 &30.67&(30.18, 31.16)& 30.62 & (30.18, 31.05) & NA\\
        13-17 &40.29&(39.77, 40.81)& 41.00 & (40.53, 41.47) & NA \\
        \hline
        \multicolumn{6}{c}{Races (missing rate: 0.47\%)} \\
        \hline
        White&77.19&(76.74, 77.63)& 76.55 & (76.14, 76.95) & 0.7\\
        Black&6.84&(6.58, 7.11)& 7.48 & (7.23, 7.73) & 0.5\\
        Other&15.97&(15.58, 16.36)& 15.98 & (15.63, 16.32) & 0.9\\
		Hispanic&13.14&(12.79, 13.50)& 13.54 & (13.22, 13.87) & 0.3\\
        \hline
        \multicolumn{6}{c}{Sex (missing rate: 0.08\%)} \\
        \hline
        Male&51.72&(51.19, 52.25)& 51.68 & (51.21, 52.16) & 0.1\\
        \hline
        \multicolumn{6}{c}{Adult Education (missing rate: 0\%)} \\
        \hline
        Less than high school &2.46&(2.30, 2.63)& 2.70 & (2.55, 2.86) & NA\\
        High school or higher&12.27&(11.93, 12.62)& 13.37 & (13.05, 13.70) & NA\\
        Some college&22.12&(21.68, 22.56)& 22.59 & (22.19, 22.98) & NA\\
        Higher than college&63.15&(62.64, 63.66)& 61.34 & (60.88, 61.80) & NA\\
		\hline
        \multicolumn{6}{c}{Insurance Type (missing rate: 1.49\%)} \\
        \hline
        Uninsured&20.36&(19.94, 20.80)& 21.26 & (20.87, 21.65) & 3.3\\
        Public only &70.90&(70.42, 71.38)& 69.61 & (69.17, 70.06) & 3.3\\
        Private only&4.08&(3.87, 4.29)& 4.12 & (3.93, 4.31) & 2.3\\
        Public and Private&4.66&(4.44, 4.89)& 5.00 & (4.79, 5.21) & 3.6\\
		\hline\hline
	\end{tabular}
	\label{tab:NSCH_VBMI}
\end{table}

\section{Discussion} 
\label{sec:discussion}

In this work, we developed a variational Bayesian inference method for approximate Bayesian inference of the high-dimensional linear mixed-effect model in the presence of missing responses. The sparsity structure of the regression coefficient vector can be modeled by a spike-and-slab prior on the entries of the regression coefficient. The computation algorithm is easy to implement, efficient, and faster than the classical Markov chain Monte Carlo samplers because of the vectorized updating formula for $\bbeta$, circumventing the need to explore the entire model selection space with exponential possibilities. The variational Bayesian inference method can be further incorporated with the sequential hierarchical regression imputation strategy for continuous data and the calibration-based imputation strategy for categorical data to improve the performance of imputing missing values, which is quite flexible and powerful.  

There are, however, some future extensions. The linear mixed-effect model only serves as a continuous approximation to the categorical model, and this step of the approximation can be rather sloppy in high dimensions. The underlying reason is that the categorical model could be potentially highly nonlinear, and the working model \eqref{eqn:LME_model} may not be rich enough to capture the nonlinearity happening inside the categorical model. It would be interesting to further expand the model structure of \eqref{eqn:LME_model} by considering a nonparametric component so that it can provide a better approximation to the black-box categorical data in practice. This step can be further optimized by coping with the variational Bayesian inference algorithm so that an easy-to-implement, computationally efficient, and sufficiently sophisticated methodology can be developed to deal with the complex missing data scheme in the contemporary world of statistics. We defer this topic to the future research direction. 

\appendix
\noindent

\section*{Appendix}

In this Appendix, we provide the step-by-step derivation of the updating formulas for the variational Bayesian inference algorithm outlined in Section \ref{sub:variational_inference_with_spike_and_slab_}. 
Recall that the objective function is given by
\[
\calL(q) = \expect_q\left[\ln\frac{p(\bY_{(\mathrm{obs})}, \bY_{(\mathrm{mis})}, \bB, \bTheta)}{q(\bY_{(\mathrm{mis})}, \bB, \bTheta)}\right],
\]
where $p(\bY_{(\mathrm{obs})}, \bY_{(\mathrm{mis})}, \bB, \bTheta)$ is given by the joint model \eqref{eqn:complete_data_likelihood}, \eqref{eqn:spike_and_slab}, and \eqref{eqn:hyperpriors}, and $q(\bY_{(\mathrm{mis})}, \bB, \bTheta)$ is the variational distribution
\begin{align*}
q(\bY_{(\mathrm{mis})}, \bB, \bTheta)\mathrm{d}\bY_{(\mathrm{mis})}\mathrm{d}\bB\mathrm{d}\bTheta
& = 
q(\bY_{(\mathrm{mis})}\mid (\widehat{\mu}_{y_{ij}}, \widehat{\sigma}_{y_{ij}}^2)_{(i, j)\in\calI_{(\mathrm{mis})}})\mathrm{d}\bY_{(\mathrm{mis})} \\
    &\quad\times
    q(\bbeta,\gamma\mid (\theta_k, \widehat{\mu}_{\beta_k}, \widehat{\sigma}_{\beta_k}^2)_{k = 1}^p)\mathrm{d}\bbeta\\
    &\quad\times
    q(\bB\mid\widehat{\bmu}_{\bb_1},\widehat{\bPsi}_1,\ldots,\widehat{\bmu}_{\bb_m}, \widehat{\bPsi}_m)\mathrm{d}\bB
    \\
    &\quad\times
    q(\sigma_e^2\mid\hat{a}_{\sigma_e^2},\hat{b}_{\sigma_e^2})\mathrm{d}\sigma_e^2
    q(\bPsi^{-1}\mid\widehat{\bPsi})\mathrm{d}\bPsi^{-1}\\
    &\quad\times
    q(\mu_0\mid\hat{\mu}_{\mu_0},\hat{\sigma}_{\mu_0}^2)\mathrm{d}\mu_0
    q(\sigma_0^2\mid\hat{a}_{\sigma_0^2},\hat{b}_{\sigma_0^2})\mathrm{d}\sigma_0^2\\
    &\quad\times
    q(w\mid\hat{a}_w,\hat{b}_w)\mathrm{d}w.
\end{align*}
The specific form of $q$ is given in \eqref{eqn:variational_distribution}. The computation of the entire objection function $\calL(q)$ is tedious and unnecessary for deriving the coordinate-ascent variational Bayesian inference updating formula for each block of the variational parameters. Instead, when focusing on the derivation of a fixed block of the variational parameter, we only need to consider the likelihood involving the latent variable and its variational distribution. Below, we discuss each updating rule separately. 

\section{Updating $\mathbf{\Psi}$} 
\label{sec:updating_bpsi}

By construction, we have
\begin{align*}
\argmax_{\widehat{\bPsi}}\calL(q)& = \argmax_{\widehat{\bPsi}}\left[ p(\widehat{\bPsi}) + \sum_{i = 1}^m\ln p(\bb_i\mid\widehat{\bPsi})\right]:=\argmax_{\widehat{\bPsi}}\Omega(\widehat{\bPsi}),
\end{align*}
where
\begin{align*}
\Omega(\widehat{\bPsi}) & = -\frac{m}{2}\ln|\widehat\bPsi|-\frac{1}{2}\sum_{i = 1}^m\expect_q[\bb_i\transpose\widehat\bPsi^{-1}\bb_i]+\frac{\nu - l - 1}{2}\ln|\widehat\bPsi|^{-1}-\frac{1}{2}\mathrm{tr}(\widehat\bPsi^{-1}\bV^{-1})\\
  &=\frac{m+\nu-l-1}{2}\ln|\bPsi|^{-1}-\frac{1}{2}\mathrm{tr}(\bPsi^{-1}(\bV^{-1}+\sum_i\expect_q[\bb_i\bb_i\transpose])).
\end{align*}
Maximizing $\Omega(\widehat{\bPsi})$ over the positive definite cone in $\mathbb{R}^{l\times l}$ yields that
\begin{align*}
\widehat{\bPsi} & = \frac{1}{m + \nu - l - 1}\left(\sum_{i = 1}^m\expect_q[\bb_i\bb_i\transpose]+\bV^{-1}\right)
\\
& = \frac{1}{m + \nu - l - 1}\left\{\sum_{i = 1}^m(\widehat{\bmu}_{\bb_i}\widehat{\bmu}_{\bb_i}\transpose + \widehat{\bPsi}_{\bb_i})+\bV^{-1}\right\}.
\end{align*}

\section{Updating variational parameters for $\bY_{(\mathrm{mis})}$} 
\label{sec:updating_missing_data}

We first fix the index $(i, j)$. This reduces to solving the problem
  \begin{align*}
  \argmax_{(\widehat{\mu}_{y_{ij}}, \widehat{\sigma}_{y_{ij}}^2)}\calL(q)
   = \argmax_{(\widehat{\mu}_{y_{ij}}, \widehat{\sigma}_{y_{ij}}^2)}\expect_q\left[\ln
   \frac{p(y_{ij}\mid\bbeta, \bb_i, \sigma_e^2)}{q(y_{ij}\mid\widehat{\mu}_{y_{ij}}, \widehat{\sigma}_{y_{ij}}^2)}
   \right].
  \end{align*}
  Observe that
  \begin{align*}
  \Omega(\widehat{\mu}_{y_{ij}}, \widehat{\sigma}_{y_{ij}}^2)
  & := \expect_q\left[\ln
   \frac{p(y_{ij}\mid\bbeta, \bb_i, \sigma_e^2)}{q(y_{ij}\mid\widehat{\mu}_{y_{ij}}, \widehat{\sigma}_{y_{ij}}^2)}
   \right]\\
   & = -\frac{1}{2}\ln(2\pi) - \frac{1}{2}\expect_q\left[\ln(\sigma_e^2)\right] - \expect_q\left[\frac{1}{2\sigma_e^2}\right]\expect_q\left[(y_{ij} - \bx_{ij}\transpose{}\bbeta - \bz_{ij}\transpose{}\bb_i)^2\right]\\
   &\quad + \frac{1}{2}\ln(2\pi) + \frac{1}{2}\ln(\widehat{\sigma}_{y_{ij}}^2) + \frac{1}{2\widehat{\sigma}_{y_{ij}}^2}\expect_q\left[(y_{ij} - \widehat{\mu}_{y_{ij}})^2\right]\\
   & = -\frac{\widehat{a}_{\sigma_e^2}}{2\widehat{b}_{\sigma_e^2}}\left[\expect_q(y_{ij}^2) - \expect_q(y_{ij})\expect_q(\bx_{ij}\transpose{}\bbeta + \bz_{ij}\transpose{}\bb_i)\right] + \frac{1}{2}\ln(\widehat{\sigma}_{y_{ij}}^2) + \text{constant}\\
   & = -\frac{\widehat{a}_{\sigma_e^2}}{2\widehat{b}_{\sigma_e^2}}\left[\widehat{\mu}_{y_{ij}}^2 + \widehat{\sigma}_{y_{ij}}^2 - \widehat{\mu}_{y_{ij}}^2
   \left(\sum_{k = 1}^px_{ijk}\theta_k\widehat{\mu}_{\beta_k} + \bz_{ij}\transpose{}\widehat{\bmu}_{\bb_i}\right)
    \right]\\
    &\quad + \frac{1}{2}\ln(\widehat{\sigma}_{y_{ij}}^2) + \text{constant}.
  \end{align*}
  Now we proceed to solve
  \begin{align*}
  \frac{\partial}{\partial\widehat{\mu}_{y_{ij}}}\Omega(\widehat{\mu}_{y_{ij}}, \widehat{\sigma}_{y_{ij}}^2)
  & = 0\quad\Longrightarrow\quad\widehat{\mu}_{y_{ij}} = \sum_{k = 1}^px_{ijk}\theta_k\widehat{\mu}_{\beta_k} + \bz_{ij}\transpose{}\widehat{\bmu}_{\bb_i},\\
  \frac{\partial}{\partial\widehat{\sigma}^2_{y_{ij}}}\Omega(\widehat{\mu}_{y_{ij}}, \widehat{\sigma}_{y_{ij}}^2)
  & = 0\quad\Longrightarrow\quad\widehat{\sigma}_{y_{ij}}^2 = \frac{\widehat{a}_{\sigma_e}^2}{\widehat{b}_{\sigma_e}^2}.
  \end{align*}


\section{Updating variational parameters for $\mathbf{\beta}$} 
\label{sec:updating_beta}


This part is the most challenging part as there does not exist a closed-form updating formula when we optimize with regard to $\btheta = [\theta_1,\ldots,\theta_p]\transpose$. Denote by $\widehat{\bmu}_\bbeta = [\widehat{\mu}_{\beta_1},\ldots,\widehat{\mu}_{\beta_p}]\transpose$ and $\widehat{\bsigma}^2_\bbeta = [\widehat{\sigma}^2_{\beta_1},\ldots,\widehat{\sigma}^2_{\beta_p}]\transpose$. First write
\begin{align*}
\Omega(\btheta, \widehat{\bmu}_\bbeta, \widehat{\bsigma}^2_\bbeta)
& = \expect_q\left[\ln \frac{p(\bY\mid\bbeta, \bB, \sigma_0^2) p(\bbeta, \bgamma\mid\bw, \mu_0, \sigma_0)\mathrm{d}\bbeta}
{q(\bbeta, \bgamma\mid\btheta, \widehat{\bmu}_\bbeta, \widehat{\bsigma}_\bbeta^2)\mathrm{d}\bbeta}\right]\\
& = \expect_q\left[\ln p(\bY\mid\bbeta, \bB, \sigma_0^2)\right] + \expect_q\left[\ln \frac{p(\bbeta, \bgamma\mid\bw, \mu_0, \sigma_0)\mathrm{d}\bbeta}
{q(\bbeta, \bgamma\mid\btheta, \widehat{\bmu}_\bbeta, \widehat{\bsigma}_\bbeta^2)\mathrm{d}\bbeta}\right].
\end{align*}
We first consider the expected value of the log-likelihood. Write
\begin{align*}
\expect_q\left[\ln p(\bY\mid\bbeta, \bB, \sigma_0^2)\right]
& = -\frac{1}{2}\left(\frac{\widehat{a}_{\sigma_e^2}}{\widehat{b}_{\sigma_e^2}}\right)\sum_{i = 1}^m\sum_{j = 1}^{n_i}\expect_q[(y_{ij} - \bx_{ij}\transpose{}\bbeta - \bz_{ij}\transpose{}\bb_i)^2] + \text{constant}.
\end{align*}
The keystone computation is the quadratic form $\expect_q[(y_{ij} - \bx_{ij}\transpose{}\bbeta - \bz_{ij}\transpose{}\bb_i)^2]$. When we focus on the parameter $\btheta, \widehat{\bmu}, \widehat{\bsigma}_\bbeta^2$, we have
\begin{align*}
\expect_q[(y_{ij} - \bx_{ij}\transpose{}\bbeta - \bz_{ij}\transpose{}\bb_i)^2]
& = \expect_q(\bbeta\transpose{}\bx_{ij}\bx_{ij}\transpose{}\bbeta) - 2\expect_q(\bbeta)\transpose{}\bx_{ij}\expect_q(y_{ij} - \bz_{ij}\transpose{}\bb_i) + \mathrm{constant}\\
& = \mathrm{tr}\left\{\bx_{ij}\bx_{ij}\transpose{}\expect_q(\bbeta\bbeta\transpose{})\right\} - 2\expect_q(\bbeta)\transpose{}\bx_{ij}\expect_q(y_{ij} - \bz_{ij}\transpose{}\bb_i)\\
&\quad + \mathrm{constant}.
\end{align*}
Let $\bTheta = \mathrm{diag}(\btheta)$ and $\bU = \mathrm{diag}(\widehat{\bmu}_\bbeta)$. 
By definition of $q(\bbeta\mid \btheta) = \sum_\bgamma q(\bbeta,\bgamma\mid\btheta)$, we have
\begin{align*}
\expect_q(\bbeta\bbeta\transpose{})
& = \bTheta(\eye - \bTheta)\bU + \bTheta\mathrm{diag}(\widehat{\bsigma}_\bbeta^2) + \bU\btheta\btheta\transpose{}\bU\transpose{}.
\end{align*}
It follows that
\begin{align*}
\mathrm{tr}\left\{\bx_{ij}\bx_{ij}\transpose{}\expect_q(\bbeta\bbeta\transpose{})\right\}
& = \left(\sum_{k = 1}^px_{ijk}\theta_k\widehat{\mu}_{\bbeta_k}\right)^2 + \sum_{k = 1}^px_{ijk}^2[\theta_k(1 - \theta_k)\widehat{\mu}_{\beta_k}^2 + \theta_k\widehat{\sigma}_{\beta_k}^2]\\
& = \left(\sum_{k = 1}^px_{ijk}\theta_k\widehat{\mu}_{\bbeta_k}\right)^2 + \sum_{k = 1}^p\theta_k x_{ijk}^2(\widehat{\mu}_{\beta_k}^2 + \widehat{\sigma}_{\beta_k}^2) - \sum_{k = 1}^p\theta_k^2x_{ijk}^2\widehat{\mu}_{\beta_k}^2\\
& = (\bx_{ij}\transpose{}\bU\btheta)^2 - \btheta\transpose{}\bU\mathrm{diag}(\bx_{ij}^2)\bU\btheta + (\widehat{\bsigma}_{\bbeta}^2 + \widehat{\bmu}_{\bbeta}^2)\transpose{}\mathrm{diag}(\bx_{ij}^2)\btheta.
\end{align*}
Therefore, we obtain:
\begin{align*}
\begin{aligned}
\expect_q[(y_{ij} - \bx_{ij}\transpose{}\bbeta - \bz_{ij}\transpose{}\bb_i)^2]
& = (\bx_{ij}\transpose{}\bU\btheta)^2 - \btheta\transpose{}\bU\mathrm{diag}(\bx_{ij}^2)\bU\btheta
 + (\widehat{\bsigma}_{\bbeta}^2 + \widehat{\bmu}_{\bbeta}^2)\transpose{}\mathrm{diag}(\bx_{ij}^2)\btheta\\
&\quad - 2\btheta\transpose{}\bU\bx_{ij}(\langle y_{ij}\rangle - \bz_{ij}\transpose{}\widehat{\bmu}_{\bb_i}).
\end{aligned}
\end{align*}
Here we set
\[
\widehat{y}_{ij} = \left\{
          \begin{aligned}
            &y_{ij},&\quad\text{if }(i, j)\notin\calI_{(\mathrm{mis})},\\
            &\widehat{\mu}_{y_{ij}},&\quad\text{if }(i, j)\in\calI_{(\mathrm{mis})}.
          \end{aligned}
          \right.
\]
Hence, we obtain
\begin{align}\label{eqn:Eq_loglikelihood}
\begin{aligned}
\expect_q\left[\ln p(\bY\mid\bbeta, \bB, \sigma_0^2)\right]
& = -\frac{1}{2}\left(\frac{\widehat{a}_{\sigma_e^2}}{\widehat{b}_{\sigma_e^2}}\right)\sum_{i = 1}^m\sum_{j = 1}^{n_i}(\bx_{ij}\transpose{}\bU\btheta)^2\\
&\quad + \frac{1}{2}\left(\frac{\widehat{a}_{\sigma_e^2}}{\widehat{b}_{\sigma_e^2}}\right)\sum_{i = 1}^m\sum_{j = 1}^{n_i}\btheta\transpose{}\bU\mathrm{diag}(\bx_{ij}^2)\bU\btheta\\
&\quad - \frac{1}{2}\left(\frac{\widehat{a}_{\sigma_e^2}}{\widehat{b}_{\sigma_e^2}}\right)\sum_{i = 1}^m\sum_{j = 1}^{n_i}  (\widehat{\bsigma}_{\bbeta}^2 + \widehat{\bmu}_{\bbeta}^2)\transpose{}\mathrm{diag}(\bx_{ij}^2)\btheta\\
&\quad + \left(\frac{\widehat{a}_{\sigma_e^2}}{\widehat{b}_{\sigma_e^2}}\right)\sum_{i = 1}^m\sum_{j = 1}^{n_i}\btheta\transpose{}\bU\bx_{ij}(\langle y_{ij}\rangle - \bz_{ij}\transpose{}\widehat{\bmu}_{\bb_i}) + \text{constant}.
\end{aligned}
\end{align}
We then compute the second term in $\Omega(\btheta, \widehat{\bmu}_\bbeta, \widehat{\bsigma}^2_\bbeta)$:
\begin{align}\label{eqn:Eq_loglikelihood_ratio_theta}
\begin{aligned}
&\expect_q\left[\ln \frac{p(\bbeta, \bgamma\mid\bw, \mu_0, \sigma_0^2)\mathrm{d}\bbeta}
{q(\bbeta, \bgamma\mid\btheta, \widehat{\bmu}_\bbeta, \widehat{\bsigma}_\bbeta^2)\mathrm{d}\bbeta}\right]\\
&\quad = \sum_{k = 1}^p\theta_k\left\{[\psi(\widehat{a}_w) - \psi(\widehat{a}_w + \widehat{b}_w)] - \frac{1}{2}\ln(2\pi) + \frac{1}{2}[\psi(\widehat{a}_{\sigma_0^2}) - \ln(\widehat{b}_{\sigma_0^2})]\right\}\\
&\quad\quad - \sum_{k = 1}^p\expect_q\left[\frac{(\beta_k - \mu_0)^2}{2\sigma_0^2}\gamma_k\right] + \sum_{k = 1}^p(1 - \theta_k)[\psi(\widehat{b}_w) - \psi(\widehat{a}_w + \widehat{b}_w)]\\
&\quad\quad - \sum_{k = 1}^p\theta_k\left[ \ln \theta_k - \frac{1}{2}\ln(2\pi) + \frac{1}{2}\ln\frac{1}{\widehat{\sigma}_{\beta_k}^2}\right] + \sum_{k = 1}^p\expect_q\left[\frac{(\beta_k - \widehat{\mu}_{\beta_k})^2}{2\widehat{\sigma}_{\beta_k}^2}\gamma_k\right]\\
&\quad\quad - \sum_{k = 1}^p(1 - \theta_k)\ln(1 - \theta_k)\\
&\quad = \sum_{k = 1}^p\theta_k\left\{[\psi(\widehat{a}_w) - \psi(\widehat{a}_w + \widehat{b}_w)] - \frac{1}{2}\ln(2\pi) + \frac{1}{2}[\psi(\widehat{a}_{\sigma_0^2}) - \ln(\widehat{b}_{\sigma_0^2})]\right\}\\
&\quad\quad - \sum_{k = 1}^p
\frac{\widehat{a}_{\sigma_0^2}}{2\widehat{b}_{\sigma_0^2}}\theta_k[(\widehat{\mu}_{\beta_k} - \widehat{\mu}_{\mu_0})^2 + \widehat{\sigma}_{\mu_0}^2 + \widehat{\sigma}_{\beta_k}^2]
 + \sum_{k = 1}^p(1 - \theta_k)[\psi(\widehat{b}_w) - \psi(\widehat{a}_w + \widehat{b}_w)]\\
&\quad\quad - \sum_{k = 1}^p\theta_k\left[ \ln \theta_k - \frac{1}{2}\ln(2\pi) + \frac{1}{2}\ln\frac{1}{\widehat{\sigma}_{\beta_k}^2}\right] + \sum_{k = 1}^p
\frac{\theta_k}{2}
 - \sum_{k = 1}^p(1 - \theta_k)\ln(1 - \theta_k)
\end{aligned}
\end{align}

\begin{itemize}
  \item[$\blacklozenge$]  We first optimize over $(\widehat{\bmu}_\bbeta,\widehat{\bsigma}^2_\bbeta)$. This step is relatively straightforward because of the closed-form solution to the stationary point. It is straightforward to obtain the derivative of $\widehat{\mu}_{\beta_k}$ using \eqref{eqn:Eq_loglikelihood} and \eqref{eqn:Eq_loglikelihood_ratio_theta}:
  \begin{align*}
  \frac{\partial\Omega}{\partial\widehat{\bmu}_{\bbeta}}
  & =
  -\frac{1}{2}\left(\frac{\widehat{a}_{\sigma_e^2}}{\widehat{b}_{\sigma_e^2}}\right)\sum_{i = 1}^m\sum_{j = 1}^{n_i}2\bTheta\bx_{ij}\bx_{ij}\bTheta\widehat{\bmu}_{\bbeta}\\
&\quad + \frac{1}{2}\left(\frac{\widehat{a}_{\sigma_e^2}}{\widehat{b}_{\sigma_e^2}}\right)\sum_{i = 1}^m\sum_{j = 1}^{n_i}2\bTheta\mathrm{diag}(\bx_{ij}^2)\bTheta\widehat{\bmu}_{\bbeta}\\
&\quad - \frac{1}{2}\left(\frac{\widehat{a}_{\sigma_e^2}}{\widehat{b}_{\sigma_e^2}}\right)\sum_{i = 1}^m\sum_{j = 1}^{n_i} 2\bTheta\mathrm{diag}(\bx_{ij}^2)\widehat{\bmu}_\bbeta\\
&\quad + \left(\frac{\widehat{a}_{\sigma_e^2}}{\widehat{b}_{\sigma_e^2}}\right)\sum_{i = 1}^m\sum_{j = 1}^{n_i}(\langle y_{ij}\rangle - \bz_{ij}\transpose{}\widehat{\bmu}_{\bb_i})\bTheta\bx_{ij}.
  \end{align*}
  Setting the gradient to $\zero$ yields
  \begin{align*}
  \widehat{\bmu}_{\bbeta}
  & = \left\{\frac{\widehat{a}_{\sigma_e^2}}{\widehat{b}_{\sigma_e^2}}\sum_{i = 1}^m\sum_{j = 1}^{n_i}[\bx_{ij}\bx_{ij}\transpose{}\bTheta + \mathrm{diag}(\bx_{ij}^2)(\eye - \bTheta)] + \frac{\widehat{a}_{\sigma_0^2}}{\widehat{b}_{\sigma_0^2}}\eye_p\right\}^{-1}\\
  &\quad\times \left[\frac{\widehat{a}_{\sigma_0^2}}{\widehat{b}_{\sigma_0^2}}\mu_0\one_p + \frac{\widehat{a}_{\sigma_e^2}}{\widehat{b}_{\sigma_e^2}}\sum_{i = 1}^m\sum_{j = 1}^{n_i}(\widehat{y}_{ij} - \bz_{ij}\transpose{}\widehat{\bmu}_{\bb_i})\bx_{ij}\right].
  \end{align*}
  We also take the gradient with regard to $\bsigma_\bbeta^2$ to obtain
  \begin{align*}
   \frac{\partial\Omega}{\partial\widehat{\bsigma}^2_{\bbeta}}
   & = - \frac{1}{2}\left(\frac{\widehat{a}_{\sigma_e^2}}{\widehat{b}_{\sigma_e^2}}\right)\sum_{i = 1}^m\sum_{j = 1}^{n_i} \mathrm{diag}(\bx_{ij}^2)\btheta - \frac{1}{2}\btheta\circ\left(\frac{1}{\widehat{\bsigma}_{\bbeta}^2}\right)
   - \frac{\widehat{a}_{\sigma_0^2}}{2\widehat{b}_{\sigma_0^2}}\btheta\circ\one_p.
  \end{align*}
  Therefore, setting the gradient to $\zero$ yields
  \[
  \widehat{\bsigma}_{\bbeta}^2 = \left[\left(\frac{\widehat{a}_{\sigma_e^2}}{\widehat{b}_{\sigma_e^2}}\right)\sum_{i = 1}^m\sum_{j = 1}^{n_i} \mathrm{diag}(\bx_{ij}^2) + \frac{\widehat{a}_{\sigma_0^2}}{2\widehat{b}_{\sigma_0^2}}\one_p\right]^{-1}
  \]

  \item[$\blacklozenge$] We next consider optimizing over $\btheta$, which is comparably more challenging. In this case, we can view $\widehat{\bmu}_\bbeta, \widehat{\bsigma}_\bbeta^2$ as constants. Invoking \eqref{eqn:Eq_loglikelihood} and \eqref{eqn:Eq_loglikelihood_ratio_theta}, we write
  \begin{align*}
  \Omega_{\widehat{\bmu}_\bbeta, \widehat{\bsigma}_\bbeta^2}(\btheta)&:= \Omega(\btheta, \widehat{\bmu}_\bbeta, \widehat{\bsigma}_\bbeta^2)\\
  & = \sum_{k = 1}^p\theta_k\left\{[\psi(\widehat{a}_w) - \psi(\widehat{a}_w + \widehat{b}_w)] + \frac{1}{2}[\psi(\widehat{a}_{\sigma_0^2}) - \ln(\widehat{b}_{\sigma_0^2})] + \frac{1}{2}\ln(\widehat{\sigma}_{\beta_k}^2)\right\} \\
  &\quad + \sum_{k = 1}^p\theta_k\left\{\frac{1}{2} - \frac{\widehat{a}_{\sigma_0^2}}{2\widehat{b}_{\sigma_0^2}}\theta_k[(\widehat{\mu}_{\beta_k} - \widehat{\mu}_{\mu_0})^2 + \widehat{\sigma}_{\mu_0}^2 + \widehat{\sigma}_{\beta_k}^2]\right\} - \sum_{k = 1}^p\theta_k\ln\theta_k\\
  &\quad + \sum_{k = 1}^p(1 - \theta_k)[\psi(\widehat{b}_w) - \psi(\widehat{a}_w + \widehat{b}_w)] - \sum_{k = 1}^p(1 - \theta_k)\ln(1 - \theta_k)\\
&\quad - \frac{1}{2}\left(\frac{\widehat{a}_{\sigma_e^2}}{\widehat{b}_{\sigma_e^2}}\right)\btheta\transpose{}\bU\left\{\sum_{i = 1}^m\sum_{j = 1}^{n_i}[\bx_{ij}\bx_{ij}\transpose{} - \mathrm{diag}(\bx_{ij} ^ 2)]\right\}\bU\btheta\\
&\quad - \frac{1}{2}\left(\frac{\widehat{a}_{\sigma_e^2}}{\widehat{b}_{\sigma_e^2}}\right)\sum_{i = 1}^m\sum_{j = 1}^{n_i} (\widehat{\bmu}_\bbeta^2 + \widehat{\bsigma}_{\bbeta}^2)\transpose{}\mathrm{diag}(\bx_{ij}^2)\btheta\\
&\quad + \left(\frac{\widehat{a}_{\sigma_e^2}}{\widehat{b}_{\sigma_e^2}}\right)\sum_{i = 1}^m\sum_{j = 1}^{n_i}\bx_{ij}(\langle y_{ij}\rangle - \bz_{ij}\transpose{}\widehat{\bmu}_{\bb_i})\bx_{ij}\transpose{}\bU\btheta.
  \end{align*}
The technical challenge in optimizing the function above over $\btheta$ is that the stationary point cannot be explicitly computed in a closed-form formula. This is due to the quadratic function
  \[
  g(\btheta) = \btheta\transpose\bDelta\bU,\quad\text{where }\bDelta = \bU\left\{\sum_{i = 1}^m\sum_{j = 1}^{n_i}[\bx_{ij}\bx_{ij}\transpose{} - \mathrm{diag}(\bx_{ij}\circ\bx_{ij})]\right\}\bU.
  \]
  We borrow the idea of \cite{huang2016variational} and consider the following linear approximation of the quadratic function $g$ at the last updated value $\btheta^{(\mathrm{old})}$:
  \begin{align*}
  g(\btheta) &\approx g(\btheta^{(\mathrm{old})}) + 2(\btheta^{(\mathrm{old})})\transpose\bDelta(\btheta - \btheta^{(\mathrm{old})})\\ 
  & = 2(\btheta^{(\mathrm{old})})\transpose{}\bDelta\btheta + g(\btheta^{(\mathrm{old})}) - 2(\btheta^{(\mathrm{old})})\transpose{}\bDelta\btheta^{(\mathrm{old})}.
  \end{align*}
  Namely, we can approximate $\Omega_{\widehat{\bmu}_\bbeta, \widehat{\bsigma}_\bbeta^2}(\btheta)$ by
  \begin{align*}
  \Omega_{\widehat{\bmu}_\bbeta, \widehat{\bsigma}_\bbeta^2}(\btheta)
  &\approx \sum_{k = 1}^p\theta_k\left\{[\psi(\widehat{a}_w) - \psi(\widehat{a}_w + \widehat{b}_w)] + \frac{1}{2}[\psi(\widehat{a}_{\sigma_0^2}) - \ln(\widehat{b}_{\sigma_0^2})] + \frac{1}{2}\ln(\widehat{\sigma}_{\beta_k}^2)\right\} \\
  &\quad + \sum_{k = 1}^p\theta_k\left\{\frac{1}{2} - \frac{\widehat{a}_{\sigma_0^2}}{2\widehat{b}_{\sigma_0^2}}\theta_k[(\widehat{\mu}_{\beta_k} - \widehat{\mu}_{\mu_0})^2 + \widehat{\sigma}_{\mu_0}^2 + \widehat{\sigma}_{\beta_k}^2]\right\} - \sum_{k = 1}^p\theta_k\ln\theta_k\\
  &\quad + \sum_{k = 1}^p(1 - \theta_k)[\psi(\widehat{b}_w) - \psi(\widehat{a}_w + \widehat{b}_w)] - \sum_{k = 1}^p(1 - \theta_k)\ln(1 - \theta_k)\\
&\quad - \left(\frac{\widehat{a}_{\sigma_e^2}}{\widehat{b}_{\sigma_e^2}}\right)(\btheta^{(\mathrm{old})})\transpose{}\bU\left\{\sum_{i = 1}^m\sum_{j = 1}^{n_i}[\bx_{ij}\bx_{ij}\transpose{} - \mathrm{diag}(\bx_{ij}\circ\bx_{ij})]\right\}\bU\btheta\\
&\quad - \frac{1}{2}\left(\frac{\widehat{a}_{\sigma_e^2}}{\widehat{b}_{\sigma_e^2}}\right)\sum_{i = 1}^m\sum_{j = 1}^{n_i} (\widehat{\bmu}_\bbeta^2 + \widehat{\bsigma}_{\bbeta}^2)\transpose{}\mathrm{diag}(\bx_{ij}\circ\bx_{ij})\btheta\\
&\quad + \left(\frac{\widehat{a}_{\sigma_e^2}}{\widehat{b}_{\sigma_e^2}}\right)\sum_{i = 1}^m\sum_{j = 1}^{n_i}\bx_{ij}(\langle y_{ij}\rangle - \bz_{ij}\transpose{}\widehat{\bmu}_{\bb_i})\bx_{ij}\transpose{}\bU\btheta + \text{constant}.
  \end{align*}
Here, $\btheta^{(\mathrm{old})}$ is the last iterate of the $\btheta$ value that can be treated as a constant. Let $\one_p = [1,\ldots,1]\transpose{}\in\mathbb{R}^p$ be the $p$-dimensional vector of all ones. Hence, we take the gradient to obtain
\begin{align*}
  \frac{\partial}{\partial\btheta}\Omega_{\widehat{\bmu}_\bbeta, \widehat{\bsigma}_\bbeta^2}(\btheta)
  & \approx
  \left\{[\psi(\widehat{a}_w) - \psi(\widehat{a}_w + \widehat{b}_w)] + \frac{1}{2}[\psi(\widehat{a}_{\sigma_0^2}) - \ln(\widehat{b}_{\sigma_0^2})]\right\}\one_p  + \frac{1}{2}\ln(\widehat{\bsigma}_{\bbeta}^2)\\
  &\quad + \frac{1}{2}\one_p  - \frac{\widehat{a}_{\sigma_0^2}}{2\widehat{b}_{\sigma_0^2}}[(\widehat{\bmu}_{\bbeta} - \widehat{\mu}_{\mu_0}\one_p)^2 + \widehat{\sigma}_{\mu_0}^2\one_p + \widehat{\bsigma}_{\bbeta}^2] - \one_p - \mathrm{logit}(\btheta)\\
  &\quad - [\psi(\widehat{b}_w) - \psi(\widehat{a}_w + \widehat{b}_w)]\one_p  + \one_p\\
&\quad - \frac{1}{2}\left(\frac{\widehat{a}_{\sigma_e^2}}{\widehat{b}_{\sigma_e^2}}\right)\sum_{i = 1}^m\sum_{j = 1}^{n_i}\bx_{ij}^2\circ(\widehat{\bmu}_{\bbeta}^2 + \widehat{\bsigma}_{\bbeta}^2)
\\
&\quad + \left(\frac{\widehat{a}_{\sigma_e^2}}{\widehat{b}_{\sigma_e^2}}\right)\sum_{i = 1}^m\sum_{j = 1}^{n_i}(\widehat{y}_{ij} - \bz_{ij}\transpose{}\widehat{\bmu}_{\bb_i})(\bx_{ij}\circ\widehat{\bmu}_{\bbeta})
\\
&\quad - \left(\frac{\widehat{a}_{\sigma_e^2}}{\widehat{b}_{\sigma_e^2}}\right)\bU\left\{\sum_{i = 1}^m\sum_{j = 1}^{n_i}[\bx_{ij}\bx_{ij}\transpose{} - \mathrm{diag}(\bx_{ij}\circ\bx_{ij})]\right\}\bU\btheta^{(\mathrm{old})}.
  \end{align*}
We now focus on the last two lines of the preceeding display. We use the updating formula for $\widehat{\bmu}_{\bbeta}$ obtained earlier to write
  \begin{align*}
  &\sum_{i = 1}^m\sum_{j = 1}^{n_i}(\widehat{y}_{ij} - \bz_{ij}\transpose{}\widehat{\bmu}_{\bb_i})\bx_{ij}\circ\widehat{\bmu}_{\bbeta} - \bU\left\{\sum_{i = 1}^m\sum_{j = 1}^{n_i}[\bx_{ij}\bx_{ij}\transpose{} - \mathrm{diag}(\bx_{ij}\circ\bx_{ij})]\right\}\bU\btheta^{(\mathrm{old})}\\
  &\quad = \sum_{i = 1}^m\sum_{j = 1}^{n_i}\bU\bx_{ij}\bx_{ij}\transpose{}\bTheta\widehat{\bmu}_{\bbeta} + \sum_{i = 1}^m\sum_{j = 1}^{n_i}\bU\mathrm{diag}(\bx_{ij}^2)(\eye - \bTheta)\widehat{\bmu}_\bbeta - \frac{\widehat{a}_{\sigma_0^2}\widehat{b}_{\sigma_e^2}}{\widehat{b}_{\sigma_0^2}\widehat{a}_{\sigma_e^2}}\widehat{\mu}_{\mu_0}\bU\eye_p\\
  &\quad\quad + \sum_{i = 1}^m\sum_{j = 1}^{n_i}\bU\mathrm{diag}(\bx_{ij}^2)\bTheta\widehat{\bmu}_{\bbeta} - \sum_{i = 1}^m\sum_{j = 1}^{n_i}\bU\bx_{ij}\bx_{ij}\transpose{}\bTheta\widehat{\bmu}_{\bbeta} + \frac{\widehat{b}_{\sigma_e^2}\widehat{a}_{\sigma_0^2}}{\widehat{a}_{\sigma_e^2}\widehat{b}_{\sigma_0^2}}\widehat{\bmu}_{\bbeta}^2\\
  &\quad = \sum_{i = 1}^m\sum_{j = 1}^{n_i}\bU\mathrm{diag}(\bx_{ij}^2) \widehat{\bmu}_\bbeta - \frac{\widehat{a}_{\sigma_0^2}\widehat{b}_{\sigma_e^2}}{\widehat{b}_{\sigma_0^2}\widehat{a}_{\sigma_e^2}}\widehat{\mu}_{\mu_0}\bU\eye_p\\
  &\quad = \sum_{i = 1}^m\sum_{j = 1}^{n_i}\bx_{ij}^2\circ \widehat{\bmu}_\bbeta^2 - \frac{\widehat{a}_{\sigma_0^2}\widehat{b}_{\sigma_e^2}}{\widehat{b}_{\sigma_0^2}\widehat{a}_{\sigma_e^2}}\widehat{\mu}_{\mu_0}\widehat{\bmu}_{\bbeta} + \frac{\widehat{b}_{\sigma_e^2}\widehat{a}_{\sigma_0^2}}{\widehat{a}_{\sigma_e^2}\widehat{b}_{\sigma_0^2}}\widehat{\bmu}_{\bbeta}^2\\
  &\quad = \frac{\widehat{b}_{\sigma_e^2}}{\widehat{a}_{\sigma_e^2}}\left(\widehat{\bmu}_{\bbeta}^2\circ\frac{1}{\widehat{\bsigma}_{\bbeta}^2} - \frac{\widehat{a}_{\sigma_0^2}}{\widehat{b}_{\sigma_0^2}}\widehat{\mu}_{\mu_0}\widehat{\bmu}_{\bbeta}\right).
  \end{align*}
This allows us to remove the last iterate $\btheta^{(\mathrm{old})}$ and hence, leads to the following approximation of the gradient
  \begin{align*}
  \frac{\partial}{\partial\btheta}\Omega_{\widehat{\bmu}_\bbeta, \widehat{\bsigma}_\bbeta^2}(\btheta)
  & \approx
  \left\{[\psi(\widehat{a}_w) - \psi(\widehat{a}_w + \widehat{b}_w)] + \frac{1}{2}[\psi(\widehat{a}_{\sigma_0^2}) - \ln(\widehat{b}_{\sigma_0^2})]\right\}\one_p  + \frac{1}{2}\ln(\widehat{\bsigma}_{\bbeta}^2)\\
  &\quad +  \frac{\widehat{a}_{\sigma_0^2}}{2\widehat{b}_{\sigma_0^2}}[(\widehat{\bmu}_{\bbeta} - \widehat{\mu}_{\mu_0}\one_p)^2 + \widehat{\sigma}_{\mu_0}^2\one_p + \widehat{\bsigma}_{\bbeta}^2] + \frac{1}{2}\one_p  - \mathrm{logit}(\btheta)\\
  &\quad - [\psi(\widehat{b}_w) - \psi(\widehat{a}_w + \widehat{b}_w)]\one_p 
   - \frac{1}{2}\left(\frac{\widehat{a}_{\sigma_e^2}}{\widehat{b}_{\sigma_e^2}}\right)\sum_{i = 1}^m\sum_{j = 1}^{n_i}\bx_{ij}^2\circ(\widehat{\bmu}_{\bbeta}^2 + \widehat{\bsigma}_{\bbeta}^2)
  \\
  &\quad
   + \widehat{\bmu}_{\bbeta}^2\circ\frac{1}{\widehat{\bsigma}_{\bbeta}^2} - \frac{\widehat{a}_{\sigma_0^2}}{\widehat{b}_{\sigma_0^2}}\widehat{\mu}_{\mu_0}\widehat{\bmu}_{\bbeta}.
  \end{align*}
  Now setting the gradient to $\zero$ gives rise to
  \begin{align*}
  \mathrm{logit}(\btheta)
  & = \left\{[\psi(\widehat{a}_w) - \psi(\widehat{a}_w + \widehat{b}_w)] + \frac{1}{2}[\psi(\widehat{a}_{\sigma_0^2}) - \ln(\widehat{b}_{\sigma_0^2})]\right\}\one_p  + \frac{1}{2}\ln(\widehat{\bsigma}_{\bbeta}^2)\\
  &\quad +  \frac{\widehat{a}_{\sigma_0^2}}{2\widehat{b}_{\sigma_0^2}}[(\widehat{\bmu}_{\bbeta} - \widehat{\mu}_{\mu_0}\one_p)^2 + \widehat{\sigma}_{\mu_0}^2\one_p + \widehat{\bsigma}_{\bbeta}^2] + \frac{1}{2}\one_p\\
  &\quad - [\psi(\widehat{b}_w) - \psi(\widehat{a}_w + \widehat{b}_w)]\one_p 
   - \frac{1}{2}\left(\frac{\widehat{a}_{\sigma_e^2}}{\widehat{b}_{\sigma_e^2}}\right)\sum_{i = 1}^m\sum_{j = 1}^{n_i}\bx_{ij}^2\circ(\widehat{\bmu}_{\bbeta}^2 + \widehat{\bsigma}_{\bbeta}^2)
  \\
  &\quad
   + \widehat{\bmu}_{\bbeta}^2\circ\frac{1}{\widehat{\bsigma}_{\bbeta}^2} - \frac{\widehat{a}_{\sigma_0^2}}{\widehat{b}_{\sigma_0^2}}\widehat{\mu}_{\mu_0}\widehat{\bmu}_{\bbeta}
  \end{align*}
  \end{itemize}

\section{Updating variational parameters for $\bB$} 
\label{sec:updating_random_effect_coef}

We now move on to the updating formula for the variational parameters for the random effect $\bB$. This part still requires some work but is significantly simpler than the previous set of parameters $(\btheta, \widehat{\bmu}_\bbeta,\widehat{\bsigma}_\bbeta^2)$. To begin with, we first compute the expected value of the quadratic form $(y_{ij} - \bx_{ij}\transpose{}\bbeta - \bz_{ij}\transpose{}\bb_i)^2$ and view the variational parameters of $y_{ij}$, $\bbeta$, and $\sigma_e^2$ as constants:
\begin{align*}
\expect_q[(y_{ij} - \bx_{ij}\transpose{}\bbeta - \bz_{ij}\transpose{}\bb_i)^2]
& = \mathrm{tr}[\bz_{ij}\bz_{ij}\transpose\expect_q(\bb_i\bb_i\transpose)]
    - 2\expect_q(\bb_i)\transpose{}\bz_{ij}(\widehat{y}_{ij} - \bx_{ij}\transpose{}\bTheta\widehat{\bmu}_{\bbeta})\\
    &\quad + \text{constant}
\\
& = \widehat{\bmu}_{\bb_i}\transpose{}\bz_{ij}\bz_{ij}\transpose{}\widehat{\bmu}_{\bb_i} + \bz_{ij}\transpose{}\widehat{\bPsi}_{\bb_i}\bz_{ij} - 2\widehat{\bmu}_{\bb_i}\transpose{}\bz_{ij}(\widehat{y}_{ij} - \bx_{ij}\transpose{}\bTheta\widehat{\bmu}_{\bbeta})\\
&\quad + \text{constant}.
\end{align*}
Hence, we can proceed to write
\begin{align*}
\Omega(\widehat{\bmu}_{\bb_i}, \widehat{\bPsi}_{\bb_i})
& = \expect_q\left[\ln\frac{p(\bY\mid\bbeta, \bB, \sigma_e^2)p(\bb_i\mid\widehat{\bPsi})}{q(\bb_i\mid\widehat{\bmu}_{\bb_i},\widehat{\bPsi}_{\bb_i})}\right]\\
& = \sum_{j = 1}^n\expect_q[\ln p(y_{ij}\mid\bbeta, \bb_i, \sigma_e^2)] + \expect_q[\ln p(\bb_i\mid\widehat{\bPsi})] - \expect_q[\ln q(\bb_i\mid\widehat{\bmu}_{\bb_i},\widehat{\bPsi}_{\bb_i})]\\
&\quad + \text{constant}\\
& = -\expect_q\left(\frac{1}{2\sigma_e^2}\right)\sum_{j = 1}^{n_i}\expect_q[(y_{ij} - \bx_{ij}\bbeta - \bz_{ij}\bb_i)^2] - \frac{1}{2}\expect_q(\bb_i\transpose{}\widehat{\bPsi}^{-1}\bb_i)\\
&\quad + \frac{1}{2}\ln\det\widehat{\bPsi}_i + \frac{1}{2}\expect_q[(\bb_i - \widehat{\bmu}_{\bb_i})\transpose{}\widehat{\bPsi}_{\bb_i}^{-1}(\bb_i - \widehat{\bmu}_{\bb_i})] + \text{constant}\\
& = -\frac{1}{2}\frac{\widehat{a}_{\sigma_e^2}}{\widehat{b}_{\sigma_e^2}}\sum_{j = 1}^{n_i}\{
\widehat{\bmu}_{\bb_i}\transpose{}\bz_{ij}\bz_{ij}\transpose{}\widehat{\bmu}_{\bb_i} + \bz_{ij}\transpose{}\widehat{\bPsi}_{\bb_i}\bz_{ij} - 2\widehat{\bmu}_{\bb_i}\transpose{}\bz_{ij}(\widehat{y}_{ij} - \bx_{ij}\transpose{}\bTheta\widehat{\bmu}_{\bbeta})
\}\\
&\quad
-\frac{1}{2}\mathrm{tr}(\hat{\bPsi}^{-1}\hat{\bPsi}_{\bb_i}) - \frac{1}{2}\widehat{\bmu}_{\bb_i}\transpose\hat{\bPsi}^{-1}\widehat{\bmu}_{\bb_i}
 + \frac{1}{2}\mathrm{tr}(\widehat{\bPsi}_{\bb_i}^{-1}\widehat{\bPsi}_{\bb_i}) + \frac{1}{2}\ln\det\widehat{\bPsi}_{\bb_i}\\
 &\quad  + \text{constant}
\end{align*}
Optimizing over $\widehat{\bmu}_{\bb_i}$, we obtain
\begin{align*}
\frac{\partial\Omega}{\partial\widehat{\bmu}_{\bb_i}}
& = -\frac{\widehat{a}_{\sigma_e^2}}{\widehat{b}_{\sigma_e^2}}\sum_{j = 1}^{n_i}[\bz_{ij}\bz_{ij}\transpose{}\widehat{\bmu}_{\bb_i} - (\widehat{y}_{ij} - \bx_{ij}\transpose{}\bTheta\widehat{\bmu}_{\bbeta})\widehat\bz_{ij}] - \widehat{\bPsi}^{-1}\widehat{\bmu}_{\bb_i} = 0\\
& \Longrightarrow\quad
\widehat{\bmu}_{\bb_i} = \left(\frac{\widehat{a}_{\sigma_e^2}}{\widehat{b}_{\sigma_e^2}}\sum_{j = 1}^{n_i}\bz_{ij}\bz_{ij}\transpose+\widehat{\bPsi}^{-1}\right)^{-1}\left[\frac{\widehat{a}_{\sigma_e^2}}{\widehat{b}_{\sigma_e^2}}\sum_{j = 1}^{n_i}(y_{ij}-\bx_{ij}\transpose\bbeta)\bz_{ij}\right].
\end{align*}
Optimizing over $\widehat{\bPsi}_{\bb_i}$ in the positive definite cone in $\mathbb{R}^{l\times l}$, we obtain
\begin{align*}
  \widehat{\bPsi}_{\bb_i} = \left(\frac{\widehat{a}_{\sigma_e^2}}{\widehat{b}_{\sigma_e^2}}\sum_{j = 1}^n\bz_{ij}\bz_{ij}\transpose+\hat{\bPsi}^{-1}\right)^{-1}.
\end{align*}


\section{Updating variational parameters for $\sigma_e^2$} 
\label{sec:updating_variational_parameters_for_error}

The updating formula for $\widehat{a}_{\sigma_e^2}$ and $\widehat{b}_{\sigma_e^2}$ involves a delicate and complete computation of the expected value of the quadratic form $\expect_q[(y_{ij} - \bx_{ij}\transpose{}\bbeta - \bz_{ij}\transpose{}\bb_i)^2]$. Write
\begin{align*}
&\expect_q[(y_{ij} - \bx_{ij}\transpose{}\bbeta - \bz_{ij}\transpose{}\bb_i)^2]\\
&\quad
= \expect_q[(y_{ij})^2] + \mathrm{tr}[\bz_{ij}\bz_{ij}\transpose{}\expect_q(\bb_i\bb_i\transpose{})]
   + \mathrm{tr}[\bx_{ij}\bx_{ij}\transpose{}\expect_q(\bbeta\bbeta\transpose{})]\\
&\quad\quad - 2\expect_q(y_{ij})[\bx_{ij}\transpose{}\expect_q(\bbeta) + \bz_{ij}\transpose{}\expect_q(\bb_i)] + 2\expect_q(\bbeta)\transpose{}\bx_{ij}\expect_q(\bb_i)\transpose{}\bz_{ij}\\
& = \langle y_{ij}^2\rangle + (\bz_{ij}\transpose{}\widehat{\bmu}_{\bb_i})^2 + \bz_{ij}\transpose{}\widehat{\bPsi}_{\bb_i}\bz_{ij} + \left(\bx_{ij}\bTheta\widehat{\bmu}_{\bbeta}\right)^2 + \bx_{ij}^2\bTheta(\eye_p - \bTheta)\widehat{\bmu}_{\bbeta}^2 + \bx_{ij}^2\bTheta\widehat{\bsigma}_{\bbeta}^2
  \\
&\quad\quad - 2\widehat{y}_{ij}[\bx_{ij}\transpose{}\bTheta\widehat{\bmu}_{\bbeta} + \bz_{ij}\transpose{}\widehat{\bmu}_{\bb_i}] + 2\widehat{\bmu}_{\bbeta}\transpose{}\bTheta\bx_{ij}\widehat{\bmu}_{\bb_i}\transpose{}\bz_{ij}.
\end{align*}
Namely, by setting
\begin{align*}
\text{SSR}
& = \sum_{i = 1}^m\sum_{j = 1}^{n_i}\expect_q[(y_{ij} - \bx_{ij}\transpose{}\bbeta - \bz_{ij}\transpose{}\bb_i)^2]\\
& = \sum_{i = 1}^m\sum_{j = 1}^{n_i}\left\{\langle y_{ij}^2\rangle + (\bz_{ij}\transpose{}\widehat{\bmu}_{\bb_i})^2 + \bz_{ij}\transpose{}\widehat{\bPsi}_{\bb_i}\bz_{ij}\right\}\\
&\quad + \sum_{i = 1}^m\sum_{j = 1}^{n_i}\left\{\left(\bx_{ij}\bTheta\widehat{\bmu}_{\bbeta}\right)^2 + \bx_{ij}^2\bTheta(\eye_p - \bTheta)\widehat{\bmu}_{\bbeta}^2 + \bx_{ij}^2\bTheta\widehat{\bsigma}_{\bbeta}^2\right\}
  \\
&\quad - 2\sum_{i = 1}^m\sum_{j = 1}^{n_i}\widehat{y}_{ij}[\bx_{ij}\transpose{}\bTheta\widehat{\bmu}_{\bbeta} + \bz_{ij}\transpose{}\widehat{\bmu}_{\bb_i}] + 2\sum_{i = 1}^m\sum_{j = 1}^{n_i}\widehat{\bmu}_{\bbeta}\transpose{}\bTheta\bx_{ij}\widehat{\bmu}_{\bb_i}\transpose{}\bz_{ij},
\end{align*}
we can write
\begin{align*}
\Omega(\widehat{a}_{\sigma_e^2}, \widehat{b}_{\sigma_e^2})
& = \expect_q\left[\ln p(\bY\mid\bbeta, \bB, \sigma_e^2)\right] + \expect_q[\ln p(\sigma_e^2\mid a_1, b_1)] - \expect_q[\ln q(\sigma_e^2\mid \widehat{a}_{\sigma_e^2}, \widehat{b}_{\sigma_e^2})]\\
& = \frac{1}{2}\sum_{i = 1}^mn_i\expect_q\left[\ln\left(\frac{1}{\sigma_e^2}\right)\right] - \frac{1}{2}\expect_q\left(\frac{1}{\sigma_e^2}\right)SSR + (a_1 - 1)\expect_q\left[\ln\left(\frac{1}{\sigma_e^2}\right)\right]\\
&\quad - b_1\expect_q\left(\frac{1}{\widehat{\sigma}_e^2}\right)
 + \ln\Gamma(\widehat{a}_{\sigma_e}^2) - \widehat{a}_{\sigma_e^2}\ln(\widehat{b}_{\sigma_e^2})\\
&\quad  - 
(\widehat{a}_{\sigma_e^2} - 1)\expect_q\left[\ln\left(\frac{1}{\sigma_e^2}\right)\right]
\widehat{b}_{\sigma_e^2}\expect_q\left(\frac{1}{\widehat{\sigma}_e^2}\right)\\
& = \left(\frac{1}{2}\sum_{i = 1}^mn_i + a_1 - \widehat{a}_{\sigma_e^2}\right)[\psi(\hat{a}_{\sigma_e^2})-\ln(\widehat{b}_{\sigma_e^2})]-\frac{\widehat{a}_{\sigma_e^2}}{\widehat{b}_{\sigma_e^2}}\left[\frac{1}{2}\mathrm{SSR}+b_1-\hat{b}_{\sigma_e^2}\right]\nonumber\\
  &\quad + \ln\Gamma(\widehat{a}_{\sigma_e^2})-\widehat{a}_{\sigma_e^2}\ln(\widehat{b}_{\sigma_e^2}).
  \end{align*}
We now optimize over $(\widehat{a}_{\sigma_e^2}, \widehat{b}_{\sigma_e^2})$ by taking the derivative of $\Omega$ and setting them to zero:
\begin{align*}
  \frac{\partial\Omega}{\partial\widehat{a}_{\sigma_e^2}}&=\left(\frac{1}{2}\sum_{i = 1}^mn_i+a_1-\widehat{a}_{\sigma_e^2}\right)\psi'(\widehat{a}_{\sigma_e^2})-\frac{1}{\widehat{b}_{\sigma_e^2}}\left(\frac{1}{2}\mathrm{SSR} + b_1-\hat{b}_{\sigma_e^2}\right),\\
  \frac{\partial\Omega}{\partial\widehat{b}_{\sigma_e^2}}& = -\left(\frac{1}{2}\sum_{i 
= 1}^mn_i+a_1-\widehat{a}_{\sigma_e^2}\right)\frac{1}{\widehat{b}_{\sigma_e^2}}+\frac{\widehat{a}_{\sigma_e^2}}{\widehat{b}_{\sigma_e^2}}\left(\frac{1}{2}\mathrm{SSR} + b_1 - \widehat{b}_{\sigma_e^2}\right),\\
  \frac{\partial\Omega}{\partial\widehat{a}_{\sigma_e^2}} & =\frac{\partial\Omega}{\partial\widehat{b}_{\sigma_e^2}}=0
  \quad
  \Longrightarrow\quad
  \widehat{a}_{\sigma_e^2} = a_1+\frac{1}{2}\sum_{i = 1}^mn_i,\quad
  \widehat{b}_{\sigma_e^2} = \frac{1}{2}\mathrm{SSR} + b_1.
\end{align*}


\section{Updating the rest of the variational parameters} 
\label{sec:updating_the_rest_of_the_variational_parameters}

The updating formulas for the rest of the variational parameters can be obtained using routine methods. Write
\begin{align*}
&\Omega(\widehat{a}_{\sigma_0^2}, \widehat{b}_{\sigma_0^2})\\
&\quad = \expect_q\left[\ln\frac{p(\bbeta,\bgamma\mid w, \mu_0, \sigma_0^2)p(\sigma_0^2)}{q(\bbeta,\bgamma\mid\btheta, \widehat{\bmu}_\bbeta, \widehat{\bsigma}_\bbeta^2)q(\sigma_0^2\mid\widehat{a}_{\sigma_0^2}, \widehat{b}_{\sigma_0^2})}\right]\\
&\quad = \frac{1}{2}\sum_{k = 1}^p\theta_k[\psi(\widehat{a}_{\sigma_0^2})-\ln(\widehat{b}_{\sigma_0^2})]\\
&\quad\quad 
-\frac{\widehat{a}_{\sigma_0^2}}{2\widehat{b}_{\sigma_0^2}}\sum_{k = 1}^p\theta_k\left\{(\widehat{\mu}^2_{\mu_0}+\widehat{\sigma}^2_{\mu_0})(1-\theta_k)+[(\widehat{\mu}_{\mu_0}-\widehat{\mu}_{\bbeta_k})^2+\widehat{\sigma}^2_{\mu_0} +\widehat{\sigma}^2_{\bbeta_k}]\theta_k\right\}\\
&\quad\quad + 2[\psi(\widehat{a}_{\sigma_0^2})-\ln(\widehat{b}_{\sigma_0^2})]-\frac{\widehat{a}_{\sigma_0^2}}{\widehat{b}_{\sigma_0^2}}-(\widehat{a}_{\sigma_0^2}+1)[\psi(\widehat{a}_{\sigma_0^2})-\ln(\widehat{b}_{\sigma_0^2})]+\widehat{a}_{\sigma_0^2}\\
  &\quad\quad+\ln\Gamma(\widehat{a}_{\sigma_0^2})-\widehat{a}_{\sigma_0^2}\ln\widehat{b}_{\sigma_0^2}
  .
\end{align*}
Setting the derivatives to zero yields
\begin{align*}
  \frac{\partial\Omega}{\partial\widehat{a}_{\sigma_0^2}}&=\frac{\partial\Omega}{\partial\widehat{b}_{\sigma_0^2}}=0
  \quad\Longrightarrow\\
  \widehat{a}_{\sigma_0^2}& = 1 + \frac{1}{2}\sum_{k = 1}^p\theta_k,\\
  \widehat{b}_{\sigma_0^2}& = 1 + \frac{1}{2}\sum_{k = 1}^p\theta_k\{(\widehat{\mu}^2_{\mu_0}+\widehat{\sigma}^2_{\mu_0})(1-\theta_k)+[(\widehat{\mu}_{\mu_0}-\widehat{\mu}_{\bbeta_k})^2+\widehat{\sigma}^2_{\mu_0}+\widehat{\sigma}^2_{\bbeta_k}]\theta_k\}.
\end{align*}
For $(\widehat{a}_w, \widehat{b}_w)$, we have
\begin{align*}
    \Omega(\widehat{a}_w,\widehat{b}_w)&
    = \expect_q\left[\ln\frac{p(\bbeta,\bgamma\mid w, \mu_0, \sigma_0^2)p(w\mid a_w, b_w)}{q(\bbeta,\bgamma\mid\btheta, \widehat{\bmu}_\bbeta, \widehat{\bsigma}_\bbeta^2)q(w\mid\widehat{a}_{w}, \widehat{b}_{w})}\right]\\
    &=\sum_{k = 1}^p\expect_q[\gamma_k]\expect_q[\ln w]+\sum_{k = 1}^p\expect_q[1-\gamma_k]\expect_q[\ln(1-w)]+(a_w-1)\expect_q[\ln w]\\
    &\quad+(b_w-1)\expect_q[\ln(1-w)]-(\widehat{a}_w-1)\expect_q[\ln w]-(\widehat{b}_w-1)\expect_q[\ln(1-w)]\\
    &\quad+\ln\Gamma(\widehat{a}_w)+\ln\Gamma(\widehat{b}_w)-\ln\Gamma(\widehat{a}_w+\widehat{b}_w)\\
    &=\sum_{k = 1}^p\theta_k[\psi(\widehat{a}_w)-\psi(\widehat{a}_w+\widehat{b}_w)]+\sum_{k = 1}^p(1-\theta_k)[\psi(\widehat{b}_w)-\psi(\widehat{a}_w+\widehat{b}_w)]\\
    &\quad + (a_w-1)[\psi(\widehat{a}_w)-\psi(\widehat{a}_w+\widehat{b}_w)] + (b_w - 1)[\psi(\widehat{b}_w)-\psi(\widehat{a}_w+\widehat{b}_w)]\\
    &\quad -(\widehat{a}_w-1)[\psi(\widehat{a}_w)-\psi(\widehat{a}_w+\widehat{b}_w)]-(\widehat{b}_w-1)[\psi(\widehat{b}_w)-\psi(\widehat{a}_w+\widehat{b}_w)]\\
    &\quad+\ln\Gamma(\widehat{a}_w) + \ln\Gamma(\widehat{b}_w)-\ln\Gamma(\widehat{a}_w+\widehat{b}_w)\\
    & = \left(\sum_{k = 1}^p\theta_k+a_w-\widehat{a}_w\right)\psi(\widehat{a}_w)\\
    &\quad - \left[\sum_{k = 1}^p\theta_k+\sum_{k = 1}^p(1-\theta_k)+a_w+b_w-\widehat{a}_w-\widehat{b}_w\right]\psi(\widehat{a}_w+\widehat{b}_w)\\
    &\quad + \left[\sum_{k = 1}^p(1-\theta_k)+b_w-\widehat{b}_w\right]\psi(\widehat{b}_w)+\ln\Gamma(\widehat{a}_w)+\ln\Gamma(\widehat{b}_w)\\
    &\quad-\ln\Gamma(\widehat{a}_w+\widehat{b}_w).
\end{align*}
Now we optimize over $\widehat{a}_w$ and $\widehat{b}_w$ by taking the derivatives and setting them to zero:
\begin{align*}
    \frac{\partial\Omega}{\partial\widehat{a}_w}&=\left(\sum_{k = 1}^p\theta_k+a_w-\widehat{a}_w\right)\psi'(\widehat{a}_w)
    -\psi'(\widehat{a}_w+\widehat{b}_w)\left(p+a_w+b_w-\widehat{a}_w-\widehat{b}_w\right),\\
    \frac{\partial\Omega}{\partial\widehat{b}_w}&=\left[\sum_{k = 1}^p(1-\theta_k)+b_w-\widehat{b}_w\right]\psi'(\widehat{b}_w)-\psi'(\widehat{a}_w+\widehat{b}_w)\left(p + a_w+b_w-\widehat{a}_w-\widehat{b}_w\right),\\
    \frac{\partial\Omega}{\partial\widehat{a}_w}&=\frac{\partial\Omega}{\partial\widehat{b}_w}=0
    \quad\Longrightarrow\quad
    \widehat{a}_w=\sum_{k = 1}^p\theta_k+a_w,\quad
    \widehat{b}_w=\sum_{k = 1}^p(1-\theta_k)+b_w.
\end{align*}
Finally, for $(\widehat{\mu}_{\mu_0}, \widehat{\sigma}_{\mu_0}^2)$, we have
\begin{align*}
  \Omega(\widehat{\mu}_{\mu_0},\widehat{\sigma}^2_{\mu_0})
  &= \expect_q\left[
  \ln\frac{p(\bbeta,\bgamma\mid w, \mu_0, \sigma_0^2)p(\mu_0)}{q(\bbeta,\bgamma\mid\btheta, \widehat{\bmu}_\bbeta, \widehat{\bsigma}_\bbeta^2)q(\mu_0\mid\widehat{\mu}_{\mu_0}, \widehat{\sigma}_{\mu_0}^2)}
  \right]\\
  &=-\frac{1}{2}\expect_q\left[\frac{1}{\sigma_0^2}\right]\sum_{k = 1}^p\expect_q[\gamma_k]\expect_q[(\bbeta_k-\mu_0)^2]-\frac{1}{2}\expect_q[\mu_0^2]\\
  &\quad+\frac{1}{2\widehat{\sigma}^2_{\mu_o}}\expect_q[(\mu_0-\widehat{\mu}_{\mu_0})^2]+\text{constant}\\
  &=-\frac{\widehat{a}_{\sigma_0^2}}{2\widehat{b}_{\sigma_0^2}}\sum_{k= 1}^p\theta_k\{(\widehat{\mu}^2_{\mu_0}+\widehat{\sigma}^2_{\mu_0})(1-\theta_k)+[(\widehat{\mu}_{\mu_0}-\widehat{\mu}_{\bbeta_k})^2+\widehat{\sigma}^2_{\mu_0}+\widehat{\sigma}^2_{\bbeta_k}]\theta_k\}\\
  &\quad
  -\frac{1}{2}(\widehat{\sigma}^2_{\mu_0}+\widehat{\mu}^2_{\mu_0})-\frac{1}{2}\ln\frac{1}{\widehat{\sigma}^2_{\mu_0}} + \text{constant}.
\end{align*}
Setting the derivative to zero, we have
\begin{align*}
  \frac{\partial\Omega}{\partial\widehat{\mu}_{\mu_0}}&=-\frac{\widehat{a}_{\sigma_0^2}}{2\widehat{b}_{\sigma_0^2}}\sum_{k = 1}^p\theta_k(\widehat{\mu}_{\mu_0}-\theta_k\widehat{\mu}_{\bbeta_k})-\widehat{\mu}_{\mu_0},\quad
  \frac{\partial\Omega}{\partial\widehat{\sigma}^2_{\mu_0}}
  =-\frac{\widehat{a}_{\sigma_0^2}}{2\widehat{b}_{\sigma_0^2}}\sum_{k = 1}^p\frac{\theta_k}{2}-\frac{1}{2}+\frac{1}{2\widehat{\sigma}^2_{\mu_0}},\\
  \frac{\partial\Omega}{\partial\widehat{\mu}_{\mu_0}}&=\frac{\partial\Omega}{\partial\widehat{\sigma}^2_{\mu_0}}=0\Rightarrow\\
  \widehat{\mu}_{\mu_0}&=\left(1+\frac{\widehat{a}_{\sigma_0^2}}{\widehat{b}_{\sigma_0^2}}\sum_{k = 1}^p\theta_k\right)^{-1}\left(\frac{\widehat{a}_{\sigma_0^2}}{\widehat{b}_{\sigma_0^2}}\sum_{k = 1}^p\theta_k^2\widehat{\mu}_{\bbeta_k}\right),\quad
  \widehat{\sigma}^2_{\mu_0}=\left(1+\frac{\widehat{a}_{\sigma_0^2}}{\widehat{b}_{\sigma_0^2}}\sum_{k = 1}^p\theta_k\right)^{-1}.
\end{align*}

\bibliographystyle{apalike} 
\bibliography{VBMI_reference}





\end{document}